\documentclass[acmsmall,screen,nonacm]{acmart}
\usepackage{url}
\usepackage{float}
\usepackage{algorithm}
\usepackage{algorithmic}
\usepackage{footnote}
\usepackage{pifont}
\usepackage{tcolorbox}
\usepackage{subfigure}
\usepackage{hyperref}
\usepackage{color}
\usepackage{booktabs}
\usepackage{multirow}
\usepackage{longtable}
\usepackage[normalem]{ulem} 
\usepackage{tikz} 
\usepackage{xcolor}
\usepackage{enumitem}
\usepackage{tabularx}
\usepackage{makecell}

\usepackage{hyperref}
\usepackage{adjustbox}  
\usepackage[edges]{forest}

\definecolor{categoryblue}{RGB}{0, 114, 189}
\definecolor{topicgreen}{RGB}{119, 172, 48}
\AtBeginDocument{%
  }

\setcopyright{acmlicensed}
\copyrightyear{2018}
\acmYear{2018}
\acmDOI{XXXXXXX.XXXXXXX}

\acmJournal{JACM}
\acmVolume{37}
\acmNumber{4}
\acmArticle{111}
\acmMonth{8}

\begin{document}

\title{A Roadmap for Modern Code Review: Challenges and Opportunities}

\author{Zezhou Yang}
\email{yangzezhou@stu.hit.edu.cn}
\author{Cuiyun Gao$^{*}$}
\email{gaocuiyun@hit.edu.cn}
\affiliation{%
  \institution{Harbin Institute of Technology, Shenzhen}
  \country{China}
}

\author{Zhaoqiang Guo$^{*}$}
\email{gzq@smail.nju.edu.cn}

\author{Zhenhao Li}
\email{ginozhenhaoli@acm.org}

\author{Kui Liu}
\email{brucekuiliu@gmail.com}

\affiliation{%
  \institution{Software Engineering Application Technology Lab, Huawei}
  \city{Hangzhou}
  \country{China}
}

\author{Xin Xia}
\email{xin.xia@acm.org}
\affiliation{%
  \institution{Zhejiang University}
  \city{Hangzhou}
  \country{China}
}
\affiliation{%
  \institution{Hangzhou High-Tech Zone (Binjiang) Institute of Blockchain and Data Security}
  \city{Hangzhou}
  \country{China}
}

\author{Yuming Zhou}
\email{zhouyuming@nju.edu.cn}
\affiliation{%
  \institution{State Key Laboratory for Novel Software Technology, Nanjing University}
  \city{Nanjing}
  \country{China}
}

\thanks{${*}$ The corresponding authors}

\renewcommand{\shortauthors}{Yang et al.}
\newcommand{\gzq}[1]{{\color{red}gzq: #1}}
\newcommand{\yzz}[1]{{\color{black} #1}}
\begin{abstract}

Over the past decade, modern code review (MCR) has been established as a cornerstone of software quality assurance and a vital channel for knowledge transfer within development teams. However, the manual inspection of increasingly complex systems remains a cognitively demanding and resource-intensive activity, often leading to significant workflow bottlenecks. This paper presents a comprehensive roadmap for the evolution of MCR, consolidating over a decade of research (2013--2025) into a unified taxonomy comprising improvement techniques, which focus on the technical optimization and automation of downstream review tasks, and understanding studies, which investigate the underlying socio-technical mechanisms and empirical phenomena of the review process. By diagnosing the current landscape through a strategic SWOT analysis, we examine the transformative impact of generative AI and identify critical gaps between burgeoning AI capabilities and industrial realities. We envision a future where MCR evolves from a human-driven task into a symbiotic partnership between developers and intelligent systems. Our roadmap charts this course by proposing three pivotal paradigm shifts, Context-Aware Proactivity, Value-Driven Evaluation, and Human-Centric Symbiosis, aiming to guide researchers and practitioners in transforming MCR into an intelligent, inclusive, and strategic asset for the AI-driven future.

\end{abstract}

\ccsdesc[500]{Software and its engineering~Software maintenance tools}

\keywords{Modern code review, software quality assurance, mining software repository, code change, code review comment}


\newcommand{\phead}[1]{\vspace{1mm} \noindent {\bf #1}}

\maketitle

\section{Introduction}

Modern code review (MCR) is a collaborative process where developers examine each other's code changes before integration into the codebase. As a cornerstone of the modern DevOps lifecycle, MCR has been widely adopted to ensure software quality \cite{2015-O-Morales, 2016-O-McIntoshKAH} and long-term maintainability \cite{2021-O-Tempero} in both industrial and open-source ecosystems \cite{2019-O-Alami, 2022-Soderberg}. Empirically, unreviewed commits are twice as likely to introduce defects compared to reviewed ones \cite{2015-O-Bavota}. Beyond identifying functional bugs \cite{2012-O-Wilkerson}, security vulnerabilities \cite{2014-O-Bosu, 2016-O-Biase}, and style violations \cite{2020-Paixao}, MCR serves as a vital channel for knowledge transfer \cite{2013-Rigby}, expertise sharing \cite{2022-Wen}, and the implicit standardization of development techniques \cite{2021-Cunha}.

Despite these multifaceted benefits, MCR remains an expensive and resource-intensive activity \cite{2002-Kelly}. The manual inspection of complex logic is cognitively demanding for reviewers and can lead to significant workflow bottlenecks, delaying feature releases and increasing turnaround time \cite{2016-O-Kononenko, 2022-Kudrjavets}. To mitigate these challenges, research has branched into two main directions: \textbf{improvement techniques} and \textbf{understanding studies}. The former focuses on automating downstream tasks to assist participants, such as change decomposition \cite{2019-Biase}, reviewer recommendation \cite{2015-Xia, 2023-Li}, and automated comment synthesis \cite{2018-Gupta, 2022-Li}. The latter aims to investigate the underlying socio-technical mechanisms, uncovering critical gaps in security assurance \cite{2024-Charoenwet-ESE} and highlighting how human factors, such as gender bias and destructive criticism, can shape review outcomes and team sustainability \cite{2019-Paul, 2022-Gunawardena}.

However, we are currently at a transformative junction. The emergence of Large Language Models (LLMs) and Generative AI has begun to shift the paradigm from passive tool support toward a potential AI-Human symbiotic partnership \cite{2023-Lu, 2024-Tufano}. While these technologies offer unprecedented proficiency in generating code refinements \cite{2024-Guo} and neutral feedback \cite{2024-Rahman}, they also introduce profound risks. Recent studies indicate that AI-assisted reviews may paradoxically increase verification overhead \cite{2025-Cihan}, fail to foster collective accountability \cite{2025-Alami-TOSEM}, and potentially lead to the deskilling of junior developers \cite{2025-Tseng}. \yzz{The gap between the escalating complexity of modern software systems and the stagnant capacity of human reviewers has reached a critical threshold. Without a structured roadmap to navigate these intelligent agents into the review loop, MCR risks becoming an unsustainable burden rather than a strategic asset.}

\yzz{To address these complexities, this paper presents a systematic roadmap synthesized from 327 primary studies published between 2013 and 2025. By anchoring technical innovation within the human-centric compass of empirical research, we reconcile burgeoning AI capabilities with industrial realities to ensure a sustainable development future. To rigorously diagnose the current landscape, we employ a SWOT (Strengths, Weaknesses, Opportunities, and Threats) analysis framework. This strategic diagnosis reveals that while LLMs provide significant ``Generative Proficiency,'' a profound Context Gap persists, as models often operate in isolation from project-specific history and architectural constraints. Furthermore, we identify a ``Metric Misalignment'' where traditional benchmarks fail to capture the actual utility and cognitive costs of AI-generated feedback, alongside critical socio-technical threats such as the erosion of collective ownership and developer deskilling.}

\yzz{Guided by these findings, we propose a vision for the next decade of MCR defined by contextual ubiquity. We envision a transition from asynchronous, web-based gatekeeping toward proactive, IDE-native collaboration. In this future, AI agents will serve as repo-aware mentors providing real-time guidance during the authoring phase, ensuring that technological ``improvement'' and empirical ``understanding'' work in a reciprocal feedback loop. To realize this vision, we outline three pivotal paradigm shifts: \textit{From Passive Assistants to Proactive Collaborators}, \textit{From Accuracy Metrics to Value-Driven Evaluation}, and \textit{From Automation to Human-Centric Symbiosis}.}

\textbf{Structure of the Paper.} The remainder of this paper is organized as follows: Section \ref{Sec-2} provides the background, defining the MCR workflow and our research taxonomy. Section \ref{Sec-3} details our bibliography retrieval and selection methodology. Section \ref{Sec-4} systematically reviews the current status of MCR research, focusing on improvement techniques and understanding studies, respectively. Section \ref{sec:roadmap} presents our roadmap, including a SWOT analysis and proposed paradigm shifts for future research. Finally, Section \ref{sec-conclusion} concludes the paper.

\section{Background}
\label{Sec-2}

Modern code review (MCR) has evolved over the years, from informal code reviews conducted by developers sitting next to each other to more formal processes involving specialized tools and dedicated teams. The rise of distributed teams and remote work has also led to the development of remote code review practices, where developers can conduct code reviews from different locations and time zones \cite{2018-Sadowski}.

\subsection{Workflow of MCR}

\begin{figure}[htbp]
    \centering
    \includegraphics[width=0.85\linewidth]{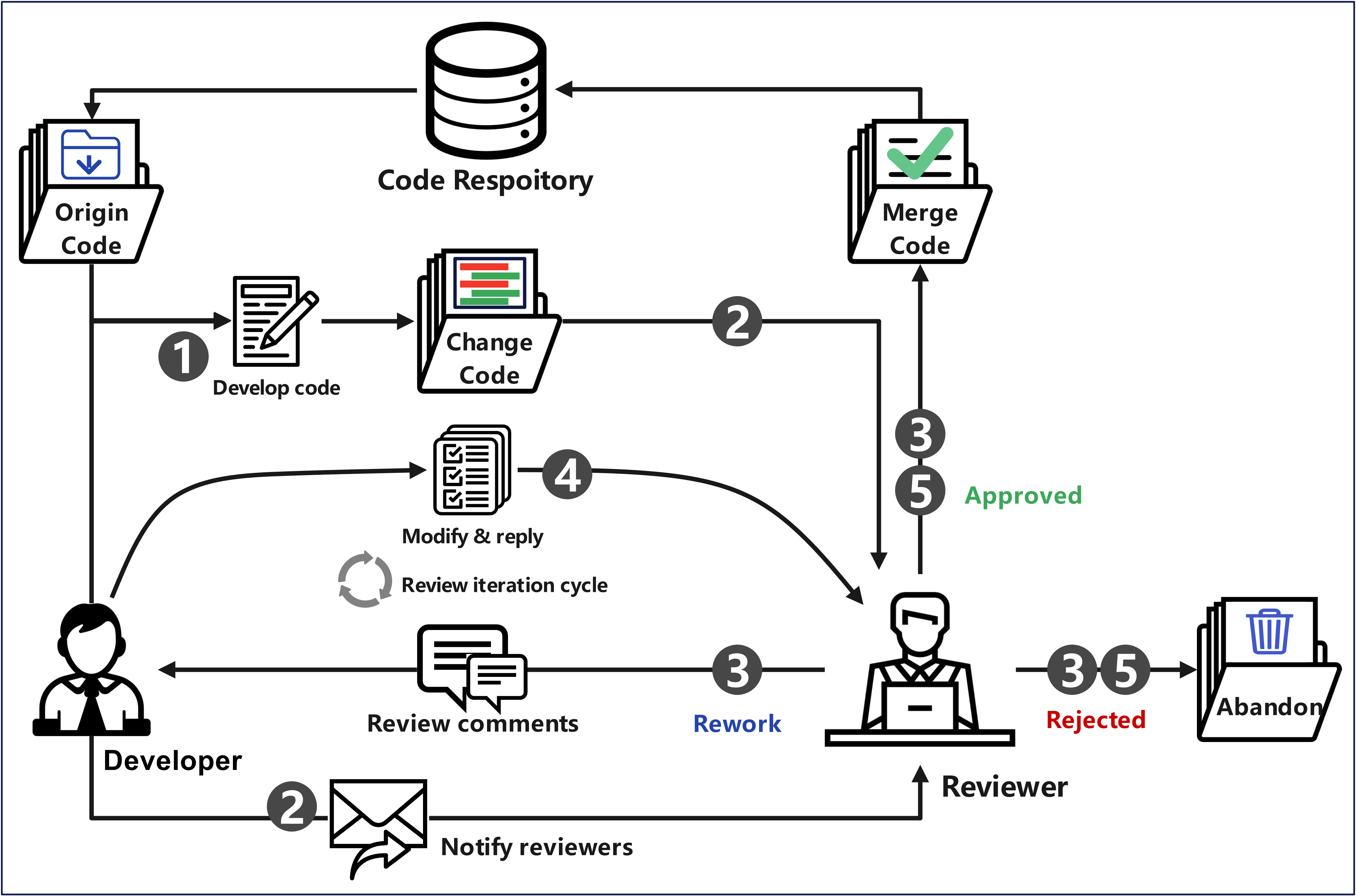}
    \caption{A general workflow of MCR.}
    \label{Figure 1}
    \Description{#}
\end{figure}

The workflow of MCR can vary depending on the organization's development process and the tools used. However, a general workflow typically involves the following steps (as shown in Figure \ref{Figure 1}):

\begin{enumerate}
    \item \textbf{Create a change.} A developer submits a code change for review. This can be done through a version control system, such as Git, or through a code review tool, such as GitHub.
    \item \textbf{Assign reviewers.} The code change is assigned to one or more reviewers, who are responsible for reviewing the code and providing feedback. Reviewers can be other developers on the team or a dedicated code review team.  
    \item \textbf{Provide comments.} The code reviewers examine the code changes and provide feedback on areas that need improvement, such as code quality, functionality, and design. The feedback can be in the form of comments, suggestions, or questions.
    \item \textbf{Modify the change and reply to comments.} The developer makes revisions to the code based on the feedback received. They can respond to comments, make code changes, or address any identified issues.    
    \item \textbf{Approve the change.} Once the code changes have been revised, the reviewer approves the changes if they are satisfied with the revisions. If there are still outstanding issues, they can request further revisions or escalate to a higher authority.
    \item \textbf{Merge the change.} Once the code changes have been approved, the developer can merge the changes into the main codebase. This makes the changes available to other developers and ensures that the codebase remains up-to-date.
\end{enumerate}

\phead{Tool Support.}
MCR is a light-weight, tool-assisted software quality assurance activity that involves the use of specialized tools that help automate the process of reviewing code changes \cite{2013-Bacchelli}. Specifically, MCR is always implemented based on code review tools. These tools make the process more effective by providing real-time feedback, automated code testing, and code analysis. Appropriate utilization of tools make it easier to identify and fix problems early in the development process, reducing the risk of introducing errors and security vulnerabilities later on \cite{2013-Rigby,2014-O-Swamidurai,2014-Beller,2018-Luxton-Reilly}. These tools can also enable developers to share code changes, track progress, and collaborate effectively on code reviews \cite{2017-Bosu,2021-Cunha,2022-Al-Rubaye}. Some of the popular code review tools used today include GitHub\footnote{https://github.com/}, GitLab\footnote{https://gitlab.com/}, Bitbucket\footnote{https://bitbucket.org/}, Gerrit\footnote{https://www.gerritcodereview.com/}, Crucible\footnote{https://www.atlassian.com/software/crucible/}, and Review Board\footnote{https://www.reviewboard.org/}.

\subsection{Research Topics of MCR}

Research in MCR can be categorized into two main directions: \textit{improvement techniques} and \textit{understanding studies}. The former focuses on enhancing the performance of specific MCR tasks through various technical approaches, while the latter encompasses comparative analyses of existing methods and exploratory investigations into specific phenomena or procedures within MCR. Given that MCR is a multi-step process involving diverse stakeholders, a comprehensive understanding from multiple perspectives is essential for developing novel methodologies. The advancement of improvement techniques drives the evolution of MCR tasks, which in turn prompts both developers and researchers to re-examine these tasks and consolidate practical experiences to enhance real-world code review processes. In summary, \textit{improvement techniques} and \textit{understanding studies} complement each other and progress together to improve MCR.

\subsubsection{Improvement Techniques}

The research on improvement techniques focuses on optimizing MCR by addressing specific downstream tasks that correspond to distinct stages of the review workflow as shown in Figure \ref{Figure 1}. These tasks are designed to assist the key entities involved in the process (i.e., the code changes, the reviewers, and the review comments), thereby enhancing the overall efficiency and quality of software development.

\begin{itemize} 
    \item \textbf{Code Change Analysis.} This task aims to assist developers and reviewers in comprehending, validating, and refining code modifications. It encompasses sub-tasks such as identifying specific change patterns \cite{2019-Shi,2022-Huang}, predicting potential defects \cite{2022-Hong-SANER,2023-Sghaier}, prioritizing review tasks \cite{2015-A-Veen,2018-Fan,2019-A-Zhao}, and improving code change quality directly \cite{2019-Ueda,2021-Tufano,2022-Thongtanunam,2025-Wang}.
    
    \item \textbf{Reviewer Recommendation.} Addressing the collaboration dynamics, this task identifies the most appropriate expert reviewer for a given change. It aims to optimize human resource allocation by comprehensively evaluating technical factors (e.g., expertise alignment \cite{2015-Xia,2016-Ouni,2023-Li}), non-technical factors (e.g., social relations and workload \cite{2017-Jiang,2019-Asthana,2025-Rigby}), and hybrid strategies that integrate these dimensions to expedite the review process \cite{2017-Xia,2022-Aryendu,2022-Rong}.
    
    \item \textbf{Review Comment Synthesis.} To augment the feedback generation process, this task aims to automate the drafting of constructive feedback. Technically, approaches are categorized into \textit{retrieval-based methods}, which extract relevant insights directly from historical data \cite{2018-Chatley, 2022-Hong-FSE}, and \textit{generation-based methods}, which synthesize novel review comments based on code changes \cite{2018-Gupta, 2020-Siow, 2024-Lu, 2025-Jaoua}.
    
    \item \textbf{Review Comment Analysis.} Focusing on the linguistic and semantic dimensions of feedback, this task evaluates comments to optimize communication and quality. It encompasses interpreting semantic content and intent \cite{2022-Ochodek, 2025-Iftikhar}, monitoring sentiment dynamics to mitigate toxicity \cite{2023-Sarker, 2024-Rahman}, and assessing feedback usefulness to prioritize actionable insights \cite{2021-Hasan, 2023-Yang}.
    
    \item \textbf{Unified Automation Frameworks.} Moving beyond isolated tasks, these frameworks integrate multiple review activities within a single system. Recent advances shift from multi-objective deep learning models \cite{2022-Tufano,2022-Li} to LLM-based paradigms, focusing on interactive multi-agent collaboration and scalable industrial deployment \cite{ren2025hydra, 2025-Sun}.
    
\end{itemize}

\subsubsection{Understanding Studies}

Different from the task-oriented focus of \textit{Improvement Techniques}, \textit{Understanding Studies} aim to investigate the underlying mechanisms, patterns, and human factors within MCR practices. These studies provide empirical evidence and user insights to guide future optimization. As MCR involves diverse stakeholders and complex interactions, researchers explore the process from the following complementary perspectives:

\begin{itemize}
    \item \textbf{Quality Assurance and Reliability.} This stream examines MCR's effectiveness in guaranteeing software quality. It covers defect detection capabilities and risk assessment \cite{2024-Yu-ESE, 2017-Singh}, security assurance gaps \cite{2016-O-Biase, 2024-Charoenwet-ESE}, and the handling of specialized artifacts ranging from documentation and test code to emerging domains like Infrastructure-as-Code (IaC) \cite{2022-Chen, 2025-Bessghaier}.

    \item \textbf{Process Efficiency and Workflow Patterns.} To optimize operational performance, this area investigates the mechanisms governing the review lifecycle. It analyzes velocity metrics such as turnaround time \cite{2022-Kudrjavets, 2025-Kansab}, evaluates the effectiveness of diverse review strategies (e.g., broadcast vs. unicast) and information presentation \cite{2017-Armstrong, 2025-Goncalves}, and identifies systemic bottlenecks prone to failures \cite{2021-Paul-ICSE, 2024-Gon}.

    \item \textbf{Human Factors and Social Interactions.} Recognizing MCR as a socio-technical process, this field explores how human cognition and interpersonal relationships shape review outcomes. It includes studies on demographic bias and social influence \cite{2014-Bosu, 2023-Sultana}, cognitive load assessment via biometrics \cite{2023-Hauser, 2025-Jetzen}, and the impact of communication styles and toxic interactions on collaboration \cite{2019-Paul, 2021-Cunha}.
    
    \item \textbf{Evolution toward Human-AI Collaboration.} Highlighting the paradigm shift driven by generative AI, this section analyzes the transition from human-centric to collaborative Human-AI systems. It covers the capabilities and performance of LLMs \cite{2024-Guo, 2025-Zeng}, integration challenges in industrial workflows \cite{2024-Davila, 2025-Cihan}, and the critical alignment of AI outputs with human oversight and trust \cite{2025-Alami-TOSEM, 2025-Salem}.
\end{itemize}

While existing improvement techniques and understanding studies have optimized the workflow of MCR, they remain predominantly rule-based or dependent on shallow learning models. Such methodologies are often confined to optimizing single, isolated tasks, failing to penetrate the intricate semantic content and complex procedural nuances inherent in the review process. The rapid evolution of LLMs now provides unprecedented opportunities for a deeper investigation and fundamental reform of modern code review \cite{ren2025hydra, sharanarthi2025adaptive, 2025-Ramesh, 2025-Sun, sharanarthi2025multi}. Consequently, a paradigm shift is essential: transitioning from fragmented auxiliary tool support to a comprehensive \textbf{AI-Human Symbiotic Partnership}. This shift motivates our proposed roadmap, which harnesses the emergent capabilities of LLMs to fundamentally reshape the MCR landscape.

\section{Overview of MCR Research}
\label{Sec-3}

\subsection{Bibliography Retrieval and Selection}
\begin{figure}[htbp]
  \centering
  \includegraphics[width=\linewidth]{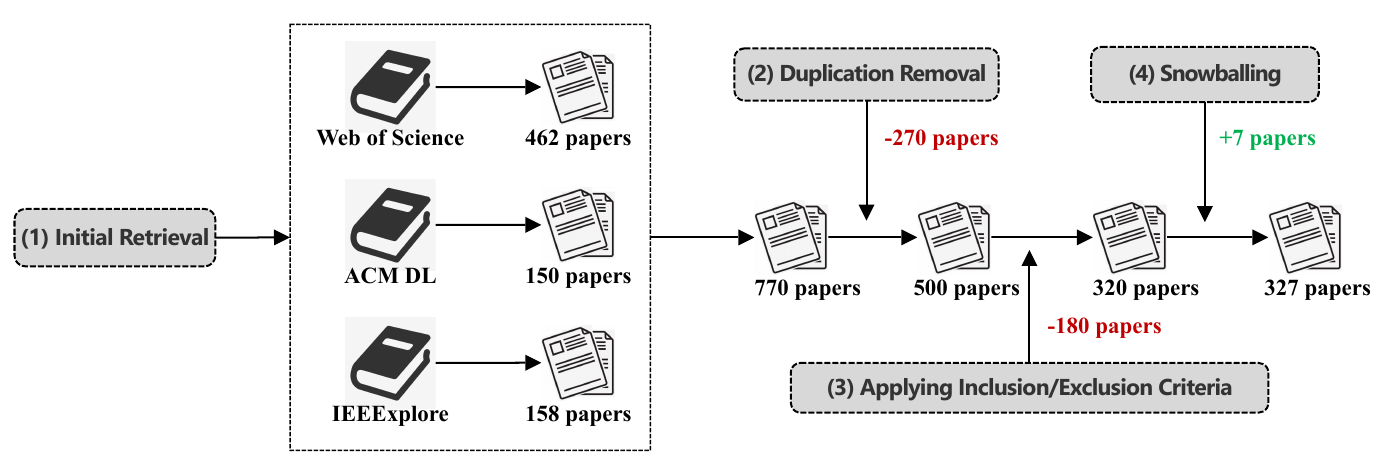}
  \caption{Bibliography retrieval and selection process map.}
  \label{Figure 2}
\end{figure}

\yzz{Following the prior guidelines provided by Petersen et al. \cite{S-Petersen}, we conduct an explicit bibliography retrieval process to collect the relevant papers in the literature, which can be reproduced by others easily.
As shown in Figure \ref{Figure 2}, the overall process consists of four steps: initial retrieval, duplication removal, inclusion \& exclusion criteria, and snowballing. The details for each step are introduced as below.}

\subsubsection{Initial Retrieval}
\yzz{The first step is to retrieve existing surveyed papers as comprehensively as possible.
To this end, the following three parameters should be taken into consideration:}
\begin{itemize}
    \item[-] \textbf{\textit{Database.}} \yzz{Similar to prior surveys \cite{S-Yang1, S-Yang2}, We select the following three online digital libraries popularly used in the field of software engineering as our retrieval databases, namely 
    Web of Science (WoS for short), 
    ACM digital library (ACM for short),
    and IEEE Xplore digital library (IEEE for short), respectively.}

    \item[-] \textbf{\textit{Query string.}} \yzz{To the best of our knowledge, ``\textit{code review}'' is the most precise term frequently used in the literature. Occasionally, a few researchers used the term ``\textit{code inspection}''. Therefore, the two groups of terms are selected as the retrieval keywords so that more relevant papers could be retrieved.
    Specifically, the query string \textit{``code review'' (Topic) or ``code inspection'' (Topic)} are used for Wos,
    the query string \textit{Title:(``code review'') OR Title:(``code inspection'')} are used for ACM,
    and the query string \textit{(``Document Title'': ``code review'') OR (``Document Title'': ``code inspection'')} are used for IEEE.}
    
    \item[-] \textbf{\textit{Time interval.}} \yzz{In the literature, the first study related to MCR is published by Bacchelli et al. \cite{2013-Bacchelli} in 2013. To conclude a complete summary of the progress of MCR, we set the retrieval time interval from January 2013 to November 2025, lasting for 13 years.}
\end{itemize}

\yzz{Following the parameters identified above, we start an initial retrieval based on the advanced function provided by each database to collect all surveyed papers whose metadata (e.g., title or topic) contains the query strings. Finally, we collect 770 surveyed papers in total, including 
462 WoS papers, 150 ACM papers, and 158 IEEE papers, respectively.}

\subsubsection{Duplication Removal}
\yzz{A common phenomenon is that an identical paper can be included in more than one online database, which leads to a huge amount of duplicated papers in our preliminary retrieval results. For the papers with multiple instances, only one distinct instance needs to be retained. To obtain these distinct papers, we manually read the title and author information of each paper to identify all duplicated ones. At the end of this step, we remove 270 duplicated papers from the initial set of papers and 500 ones are eventually retained.}

\begin{figure}[htbp]
  \centering
  \subfigure[Number of publications per year.]{
    \includegraphics[width=0.47\linewidth]{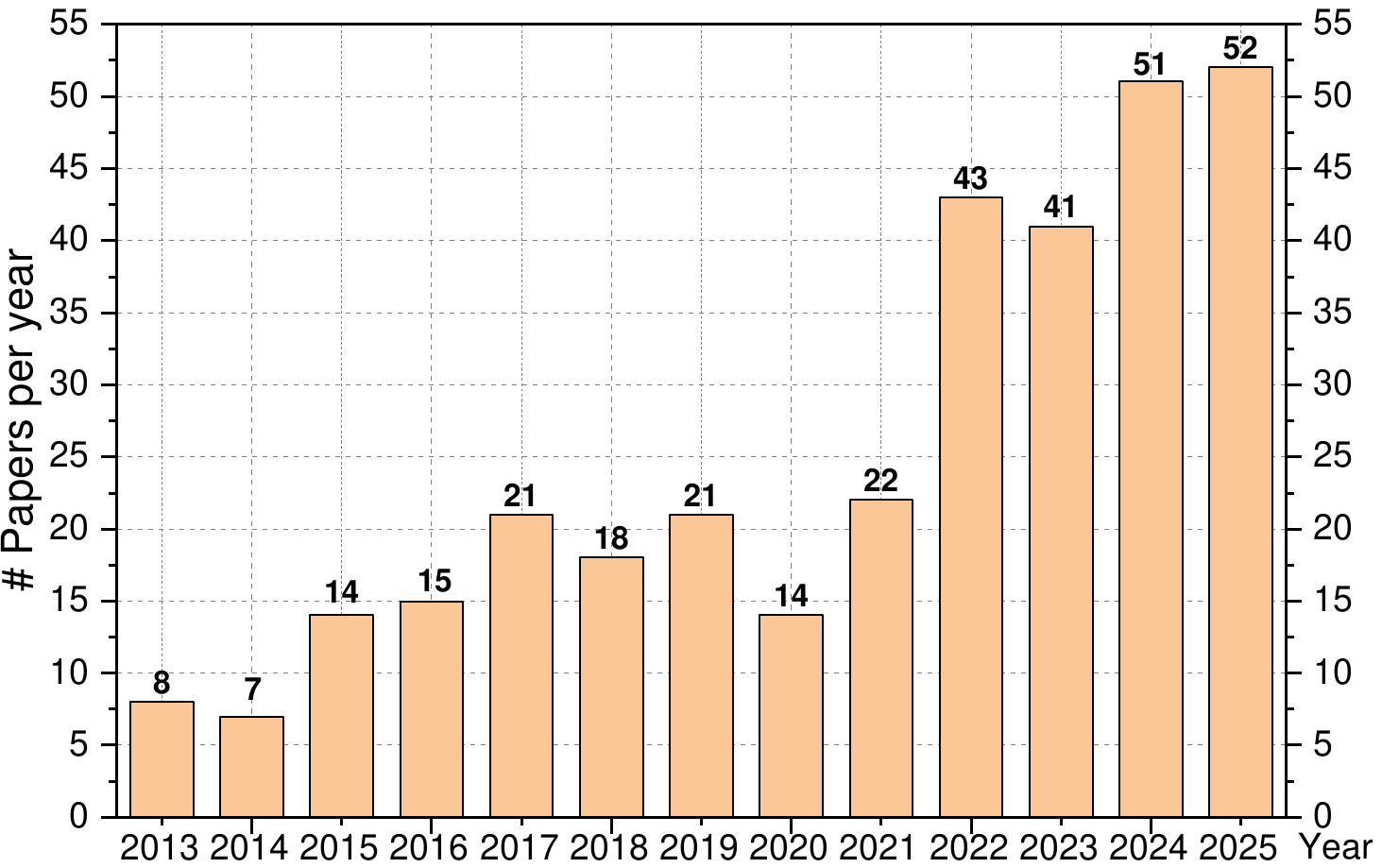}
    \label{Figure 3-1}
  }
  \quad
  \subfigure[Cumulative number of publications per year.]{
    \includegraphics[width=0.47\linewidth]{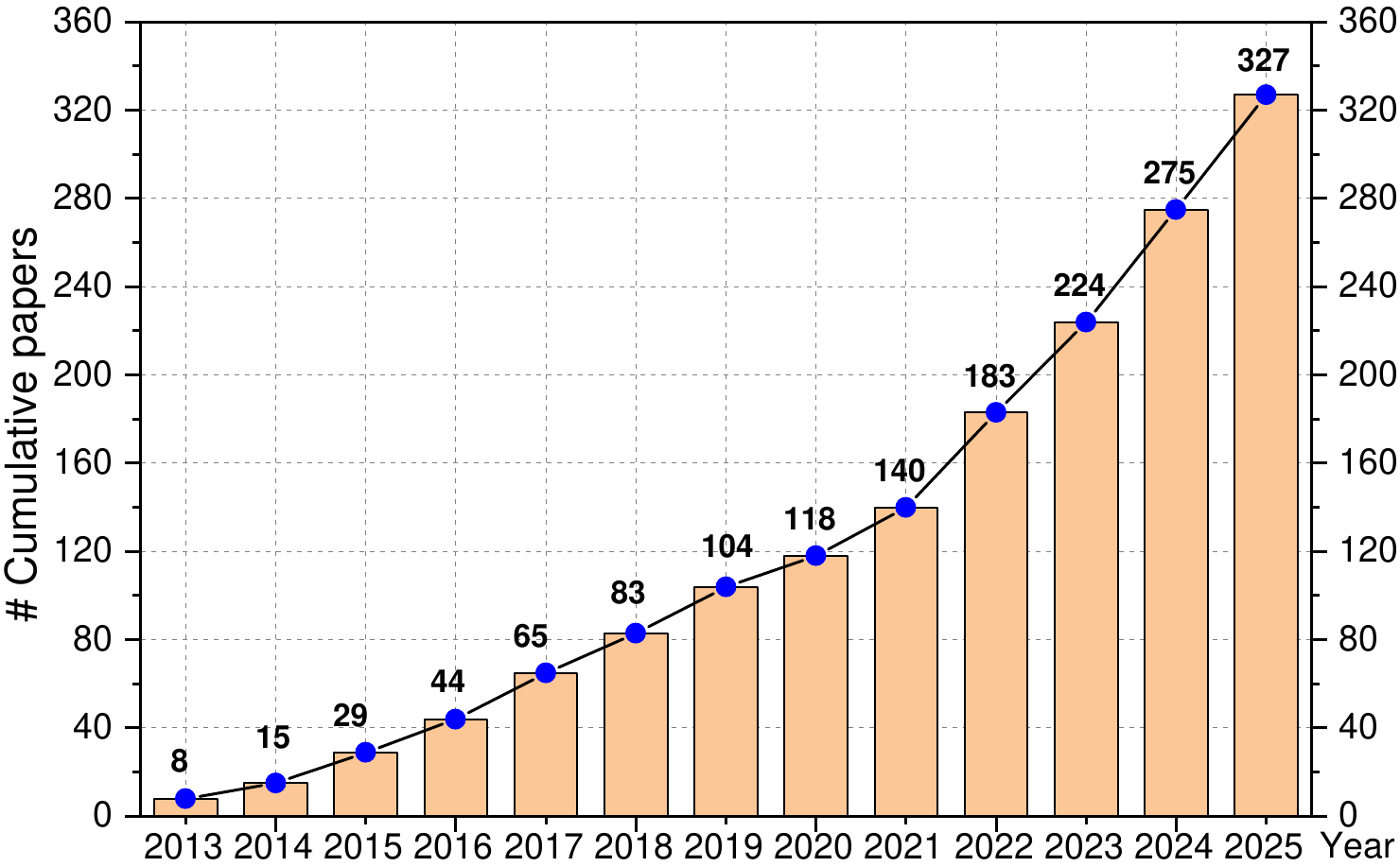}
    \label{Figure 3-2}
  }
  \caption{Publication trends of surveyed papers between 2013 and 2025.}
  \label{Figure 3}
\end{figure}

\subsubsection{\yzz{Inclusion and Exclusion Criteria}}
\yzz{Note that, a large number of papers with uneven quality or irrelevant topics can be searched under the given retrieval conditions. In order to determine the accuracy of the research papers, the following criteria are designed to select the qualified papers and filter out the irrelevant ones.}
\begin{itemize}
    \item[\ding{52}] \yzz{Studies must be written in English.}
    \item[\ding{52}] \yzz{Studies must be related to the topic of MCR.}
    \item[\ding{52}] \yzz{The length of studies must not be less than five pages.}
    \item[\ding{56}] \yzz{Studies whose research objective is not to improve MCR are dropped.}
    \item[\ding{56}] \yzz{Books, book chapters, or technical reports (most of which have been published as articles) are dropped.}
    \item[\ding{56}] \yzz{If a conference paper has an extended journal version, only the journal version is included.}
\end{itemize}

\yzz{We conduct a manual review process in this step. Specifically, we read the abstract of each paper initially to judge whether the paper fits the research topic of our survey. As a result, 180 papers are excluded after this step, and 320 surveyed studies that met our criteria are thus considered.}

\subsubsection{Snowballing}
\yzz{The final step is snowballing manually, which helps us gather other related works not included in the previous retrieval process. More specifically, we manually check the references of the selected papers to avoid missing some relevant papers whose titles do not contain the preset query strings. According to our expertise in this research domain, we manually add seven more related works that are missed by our systematic retrieval process. As a consequence, we finally identify 327 relevant surveyed papers.}

\subsection{Distribution of Surveyed Studies}
\label{sec:distribution}

\yzz{To better understand the landscape of MCR research, we analyze the 327 identified papers from three perspectives: publication trends, venue distribution, and contribution types. The detailed statistics are presented below.}

\subsubsection{Publication Trends}
\yzz{Figure \ref{Figure 3} illustrates the publication timeline of the surveyed studies from 2013 to 2025, showing both the annual number (Figure \ref{Figure 3-1}) and the cumulative number (Figure \ref{Figure 3-2}). Overall, two significant observations can be made.}

\yzz{First, there is a clear upward trend in the number of MCR papers published annually, indicating growing attention from both researchers and practitioners. Specifically, the research timeline can be categorized into three stages: the \textit{initial stage} (2013-2014, with fewer than 10 papers per year), the \textit{steady growth stage} (2015-2021, with 14$\sim$22 papers per year), and the \textit{rapid growth stage} (2022-2025, with over 40 papers per year). Notably, while previous surveys only cover literature up to 2021 \cite{2021-S-Cetin, 2021-S-Davila, 2021-S-Dong, 2021-S-Wang, 2023-Badampudi}, a dramatic increase in the volume of new studies has taken place in the years following. This phenomenon suggests that prior survey articles may have had a positive guiding effect, stimulating further academic research.}

\begin{table}[htbp]
  \centering
  \caption{The Number of Relevant Studies in Different Publication Venues. Software Engineering is the main domain keyword (bold font) of the corresponding publication venues.}
  \label{Table 2}
  \resizebox{\linewidth}{!}{
    \begin{tabular}{llr}
    \toprule
    \textbf{Acronym} & \textbf{Full name} & \textbf{Studies} \\
    \midrule
    ICSE     & ACM/IEEE International Conference on \textbf{Software Engineering}                                & 35 \\
    ESEC/FSE & ACM SIGSOFT Symposium on the Foundation of \textbf{Software Engineering}                 & 20 \\
             & European \textbf{Software Engineering} Conference                                                 &    \\
    MSR      & IEEE Working Conference on Mining Software Repositories                                           & 17 \\
    ICSME    & IEEE International Conference on Software Maintenance and Evolution                               & 16  \\
    SANER    & IEEE International Conference on Software Analysis, Evolution and Reengineering                   & 15 \\
    ASE      & IEEE/ACM International Conference on Automated \textbf{Software Engineering}                            & 10  \\
    ICPC     & IEEE International Conferences on Program Comprehensive                                           & 7  \\
    ESEM     & ACM/IEEE International Symposium on Empirical \textbf{Software Engineering} and Measurement       & 6  \\
    SCAM     & IEEE International Working Conference on Source Code Analysis and Manipulation                    & 5  \\
    EASE     & International Conference on Evaluation and Assessment in \textbf{Software Engineering}                                        & 4  \\
    APSEC    & Asia-Pacific \textbf{Software Engineering} Conference                                             & 4  \\
    ISSRE    & IEEE International Symposium on Software Reliability Engineering                                  & 3  \\
    COMPSAC  & IEEE International Computer Software and Applications Conference                                  & 2  \\
    QRS      & IEEE International Conference on Software Quality, Reliability, and Security                       & 2  \\
    ICSSP      &  International Conference on Software and System Process                       & 2  \\
    PROMISE  & International Conference on Predictive Models and Data Analytics in \textbf{Software Engineering} & 2  \\
    Others   & -                                                                                                 & 63 \\
    \midrule
    ESE     & \multicolumn{1}{l}{Empirical \textbf{Software Engineering}}                                       & 33 \\
    TSE      & \multicolumn{1}{l}{IEEE Transactions on \textbf{Software Engineering}}                            & 13  \\
    IST      & \multicolumn{1}{l}{Information and Software Technology}                                           & 14 \\
    TOSEM & \multicolumn{1}{l}{ACM Transactions on \textbf{Software Engineering} and Methodology}                                           & 12 \\
    HCI      & \multicolumn{1}{l}{Human-Computer Interaction}                                              & 6 \\   
    JSS      & \multicolumn{1}{l}{Journal of Systems and Software}                                               & 5  \\
    Others   &  -                                                                                                & 31 \\
    \bottomrule
    \end{tabular}%
  }
  \label{tab:addlabel}%
\end{table}%

\begin{figure}[htbp]
  \centering
  \subfigure[Number of papers published in different conferences.]{
    \includegraphics[width=0.48\linewidth]{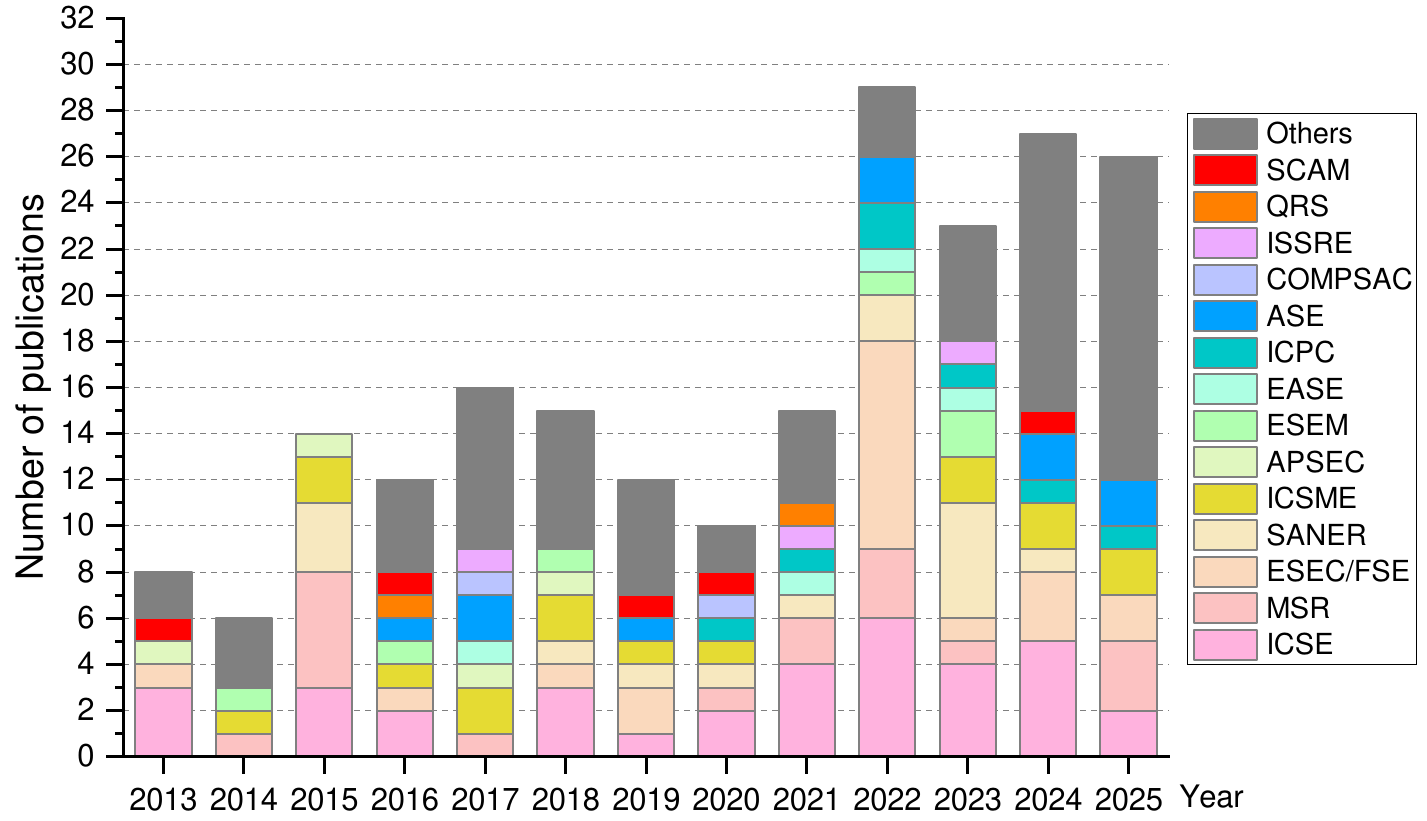}
    \label{Figure 4-1}
  }
  \quad
  \subfigure[Number of papers published in different journals.]{
    \includegraphics[width=0.46\linewidth]{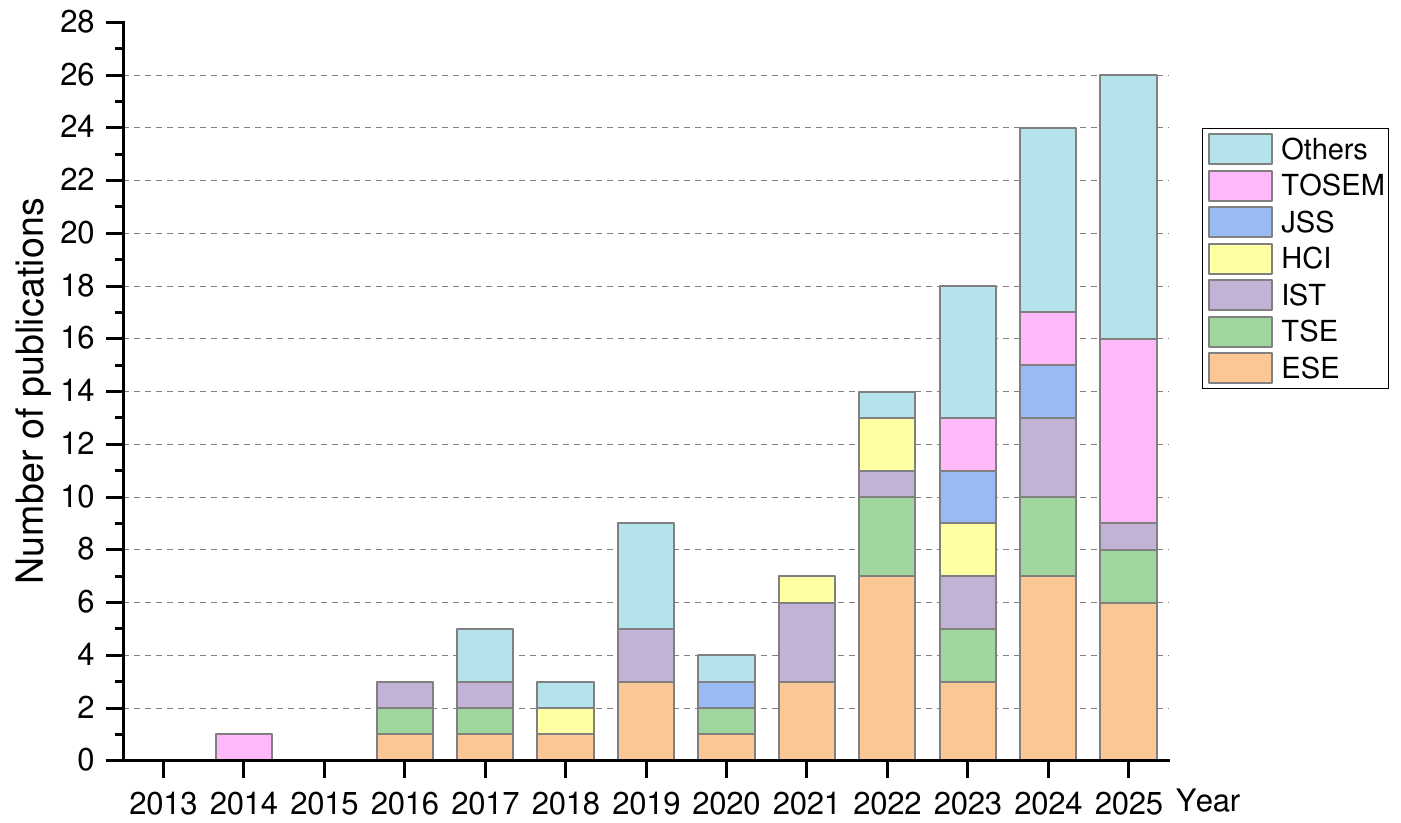}
    \label{Figure 4-2}
  }
  \caption{Publication trends for conferences (a) and journals (b) between 2013 and 2025.}
  \label{Figure 4}
\end{figure}

\yzz{Second, in terms of cumulative growth, 209 new papers have been published in the last five years (since 2021), accounting for approximately 64\% of all surveyed studies. The trajectory suggests that improving modern code review is becoming a central topic in software engineering, and we anticipate a continued increase in related studies. Overall, the publication trend demonstrates a fluctuating but persistent rise.}

\subsubsection{Venue Distribution}
\yzz{Table \ref{Table 2} lists the distribution of the surveyed papers across various publication venues, including conference proceedings and journals. We rank the venues in descending order based on the number of published papers. Venues with fewer than two papers are grouped under ``Others''. Based on Table \ref{Table 2} and Figure \ref{Figure 4}, we observe the following:}

\begin{table}[htbp]
  \centering
  \caption{Definitions and Distribution of Research Contributions in the MCR Literature.}
  \label{Table 3}
  \resizebox{\linewidth}{!}{
    \begin{tabular}{p{4.2cm}p{9.5cm}r}
    \toprule
    \textbf{Major Direction} & \textbf{Definition and Scope} & \textbf{\# (\%)} \\
    \midrule
    \textbf{Improvement Techniques} & 
    Focuses on the development of \textbf{new methodologies, algorithms, and practical tools} (e.g., plugins, frameworks) designed to automate or optimize specific MCR tasks and enhance review efficiency. & 153 (46.8\%) \\
    \midrule
    \textbf{Understanding Studies} & 
    Focuses on \textbf{empirical investigations and user studies} that utilize quantitative data mining or qualitative analysis (e.g., surveys, interviews) to uncover process patterns, human factors, and systemic bottlenecks. & 174 (53.2\%) \\
    \bottomrule
    \end{tabular}%
  }
\end{table}

\yzz{First, conference proceedings are the dominant outlet for MCR research. Specifically, conference papers account for 65\% (213 papers) of the total, while only 35\% (114 papers) appear in journals. This preference likely stems from the faster turnaround time of conferences compared to the rigorous and lengthy review cycles of journals, allowing researchers to share timely findings.}

\yzz{Second, the research is distributed across a wide range of venues (68 in total). Among journals, 20 distinct venues are identified, with \textit{Empirical Software Engineering} (ESE) being the top choice, hosting a significant portion of the journal articles. Among conferences, 48 venues are identified, with top-tier conferences such as ICSE, ESEC/FSE, and MSR accounting for the majority (150/213) of the conference papers. ICSE stands out as the most popular venue, featuring the highest number of studies.}

\yzz{Third, while the papers are scattered, the domain of \textbf{Software Engineering} remains the core focus, as evidenced by the venue titles (highlighted in bold in Table \ref{Table 2}). This keyword appears in nine top-tier venues. Other recurring keywords include ``Software Testing,'' ``Analysis,'' and ``Evolution,'' reflecting the interdisciplinary nature of MCR studies. Consequently, we recommend that researchers targeting high impact consider submitting to top-tier venues like ICSE or ESE.}

\subsubsection{Types of Contribution}
\label{sec:contributions}

\begin{figure*}[htbp]
\centering
\begin{adjustbox}{width=\textwidth, totalheight=\textheight, keepaspectratio}
\begin{forest}
  for tree={
    font=\small\sffamily,
    grow'=0,
    forked edge,
    anchor=west,
    child anchor=west,
    parent anchor=east,
    draw,
    rounded corners=2pt,
    fill=white,
    align=center, 
    base=middle,  
    l sep=8mm,     
    s sep=1.0mm,   
    edge={line width=0.9pt, draw=gray!70},
    inner sep=3pt,
  },
  where level=0{fill=gray!20, font=\bfseries\small}{},
  where level=1{fill=blue!8, font=\bfseries\small, l sep=6mm}{},
  where level=2{fill=gray!3, font=\bfseries\small, l sep=6mm}{},
  where level=3{
    draw=none, 
    fill=none, 
    font=\footnotesize, 
    text width=8cm, 
    align=left, 
    inner sep=0.5pt
  }{}, 
  [MCR\\Research
    [Improvement\\Techniques\\(Sec 4.1)
      [Code Change\\Analysis 
        [Change Decomposition (6) \cite{2015-Tao, 2015-Barnett-ICSE, 2018-Freire, 2019-Guo, 2019-Wang-ASE, 2019-Biase}]
        [Change Understanding (15) \cite{2013-He, 2021-Uchoa-MSR, 2018-Wen, 2019-Hanam, 2017-Menarini, 2014-Yamauchi} \\ \cite{2021-Wang-IST, 2024-Heander, 2022-Kanda, 2017-Wang, 2021-Balci, 2023-Fregnan, 2015-Tymchuk, 2024-KrauseGlau, 2024-Unterkalmsteiner}]
        [Specific Change Identification (17) \cite{2013-A-Meng, 2015-Zhang, 2018-Fish, 2020-Ayinala, 2022-Huang} \\ \cite{2018-Ram, 2019-Wang, 2019-Shi, 2024-Willenbring, 2024-Hong, 2023-Afzali, 2023-Imtiaz, 2023-Sawant, 2017-Chen, 2017-Ge, 2021-Brito, sieber2025classification}]
        [Defective Change Prediction (13) \cite{2016-Soltanifar, 2017-Lal, 2017-Madera, 2019-Sodhi} \\ \cite{2021-Kapur, 2021-Liu, 2023-Sghaier, 2022-Hong-SANER, 2024-Morikawa, 2024-choudhury-investigation, 2022-Chen, takieldeen2025unlocking, 2025-Peng}]
        [Change Prioritization (12) \cite{2015-O-Gousios, 2013-Aman, 2015-A-Veen, 2019-A-Zhao, 2018-Fan, 2021-Saini} \\ \cite{2022-Islam, 2022-Huang-ESE, 2024-Chouchen, 2024-Olewicki, 2024-Yang, 2025-Yang}]
        [Change Improvement (9) \\ \cite{2013-Balachandran, 2019-Ueda, 2021-Tufano, 2022-Thongtanunam, 2023-Lin, 2023-Pornprasit, 2025-Wang, 2024-FrommgenA24, 2024-Guo}]
      ]
      [Reviewer\\Recommendation 
        [Technical Factors (12) \cite{2013-Balachandran-ICSE, 2015-Thongtanunam-SANER, 2015-Xia, 2023-Li, 2016-Ouni, 2016-Rahman-ICSE} \\ \cite{2016-Zanjani, 2016-Hannebauer, 2014-Mishra, 2020-Sadman, 2021-Tecimer, 2024-Rong}]
        [Non-Technical Factors (9) \\ \cite{2016-Yu, 2017-Jiang, 2023-Zhang, 2019-Asthana, 2020-Rebai, 2020-Kovalenko, 2023-Ahmed, 2024-Hajari, 2025-Rigby}]
        [Hybrid Approaches (9) \\ \cite{2017-Xia, 2022-Pandya, 2022-Aryendu, 2022-Kong, 2022-Rong, 2024-Qiao, 2023-Liu, 2019-Suluun, 2021-Sulun}]
      ]
      [Review Comment\\Synthesis
        [Retrieval-based methods (5)  \cite{2018-Chatley, 2022-Hong-FSE, 2023-Shuvo, 2024-Kartal, ochodek2025acora}]
        [Generation-based methods (9) \\ \cite{2018-Gupta, 2020-Siow, 2022-Rahman, 2024-Lu, 2025-Jaoua, 2025-Zhang, 2025-Li, 2025-Lin, 2025-Liu}]
      ]
      [Review Comment\\Analysis
        [Content and Intent Analysis (4) \cite{2024-Kachanov, 2022-Ochodek, 2025-Iftikhar, petrova2025building}]
        [Sentiment and Tone Analysis (6) \cite{2017-Ahmed, 2020-Egelman, 2023-Sarker, 2022-Qiu, 2024-Rahman, 2020-Vrzakova}]
        [Quality and Usefulness Assessment (7) \\ \cite{2014-Pangsakulyanont, 2017-Rahman, 2021-Hasan, 2017-Ebert, 2023-Yang, 2023-Ahmed, 2025-Ahmed}]
      ]
      [Unified Automation\\Frameworks
        [Deep learning-based methods (9) \\ \cite{2022-Tufano, 2022-Li, 2023-Yin, 2024-Cao, 2024-Chouchen-TOSEM, 2024-Vijayvergiya, 2024-Xuan, 2024-Sghaier, 2025-Sghaier}]
        [LLM-based methods (11) \cite{2023-Lu, 2024-Yu-TOSEM, 2024-Almeida, 2024-chowdhury, 2024-Khelifi} \\ \cite{2024-Tang, ren2025hydra, sharanarthi2025adaptive, 2025-Ramesh, 2025-Sun, sharanarthi2025multi}]
      ]
    ]
    [Understanding\\Studies\\(Sec 4.2)
      [Quality Assurance\\and Reliability
        [Defect and Risk Detection (9) \\ \cite{2024-Yu-ESE, 2015-Thongtanunam, 2017-Izquierdo-Cortazar17, 2015-Panichella, 2017-Singh, 2023-Gunawardena, 2023-Mehrpour, 2016-Hentschel, 2018-Ueda}]
        [Security Review Effectiveness (9) \\ \cite{2023-Yu, 2013-Bacchelli, 2016-O-Biase, 2024-Charoenwet-ESE, 2023-Alfadel,2024-Charoenwet-ISSTA, 2018-Wang, 2024-AndreiCristian, 2022-Braz-FSE}]
        [Reviewing Specialized Artifacts (20) \cite{2023-Nejati, 2024-Rokem, 2022-Rao, 2018-Spadini, 2019-Spadini} \\ \cite{2018-An, 2022-AlOmar, 2020-Paixao, 2021-AlOmar, 2020-Han, 2022-Han, 2024-Tuna, 2025-Oliveira, 2025-Kim, 2025-Coelho, 2025-AlOmar, 2023-Bauser, 2022-Eisty, 2025-Bessghaier, ludwig2025enhancing}]
      ]
      [Process Efficiency and\\Workflow Patterns
        [Code Velocity and Turnaround Time (12) \cite{2024-Kudrjavets, 2022-Kudrjavets, 2023-Kudrjavets} \\ \cite{2018-Santos, 2019-Paixao, 2016-Baysal, 2022-Chen-FSE, 2022-Jiang, 2023-Chouchen-ESE-L, 2022-Al-Rubaye, 2025-Kansab, 2015-Bird}]
        [Review Strategies and Taxonomy (14) \cite{2017-Armstrong, 2016-Baum-ICSSP, 2018-Luxton-Reilly} \\ \cite{2025-Koitz-Hristov, 2022-Braz, 2022-Goncalves-ESE, 2022-Fregnan-FSE-order, 2014-Beller, 2025-Davila, 2022-Fregnan-ESE-evoluation, 2020-Panichella, 2016-Baum-QRS, 2022-Fregnan-ESE-classify, 2023-Turzo}]
        [Information Needs and Materials (22) \cite{2018-Ebert, 2018-Pascarella, 2019-Baum, 2025-Goncalves, 2019-Belgamo, 2015-Rong} \\ \cite{2019-Hirao-FSE, 2021-Wang-ESE, 2024-Zhang, 2019-Hirao, 2022-Fu, 2017-Li, 2018-Zanaty, 2021-Ebert, 2025-Chen, 2015-A-Bosu, 2017-Baum-PROFES, 2017-Baum, 2016-Kononenko, 2024-Turzo, 2024-Martin-ASE, 2025-Widyasari}]
        [Process Bottlenecks and Failures (9) \\ \cite{2017-Thongtanunam, 2019-Wang-IST, 2021-Ferreira, 2022-Shan, 2021-Paul-ICSE, 2024-Gon, 2023-Mukhtarov, 2024-Huang-JSS, 2022-Soderberg}]
      ]
      [Human Factors and\\Social Interaction
        [Demographic and Social Influence (17) \cite{2013-Baysal, 2014-Bosu, 2017-Murakami, 2023-Murphy-Hill} \\ \cite{2023-Sultana, 2019-Paul, 2024-Murphy-Hill, 2014-Meneely, 2014-Rigby, 2015-Kononenko, 2021-Wang, 2019-Ruangwan, 2022-Murphy-Hill-TSE, martins2026tokenized, 2016-Baum, 2017-Macleod, 2018-German}]
        [Biometrics and Cognitive Factors (14) \cite{2023-Hauser, 2025-Hauser, 2024-Yabesi, 2025-Wiese} \\ \cite{2021-Hijazi, 2023-Hijazi, 2023-Saranpaa, 2025-Jetzen, 2024-Middleton, 2025-Case, 2025-Tseng, 2025-Beattie, 2025-Lee, 2025-Duong}]
        [Communication and Collaboration (13) \cite{2019-Asri, 2022-Gunawardena, 2021-Chouchen} \\ \cite{2023-Sarker, 2024-Lee, 2020-Spadini, 2016-Kalyan, 2016-Hirao, 2022-Hirao, 2024-Goncalves, 2022-Goncalves, 2023-Goncalves, 2023-Ciancarini}]
        [Benefits of MCR (14) \cite{2013-Rigby, 2017-Bosu, 2021-Cunha, 2021-Sri-Iesaranusorn} \\ \cite{2022-Wen, 2023-Iftikhar, 2020-Caulo, 2025-Dorner, 2018-Sadowski, 2021-Cunha-SBQS, 2022-Li-ICSA, 2023-Petrovic, ismail2024systematic, 2025-Ardic}]
      ]
      [Evolution toward\\Human-AI \\Collaboration
        [AI Capabilities and Model Performance (6) \\ \cite{2023-Zhou, 2025-Fan, gogri2025transformer, 2025-Zeng, 2024-Tufano, 2024-Pornprasit}]
        [Integration and Practical Challenges (9)\\ \cite{2020-Wessel, 2022-Wessel, 2024-Davila, 2025-Cihan, 2023-Crandall, 2020-Wessel-SBES, 2022-Kim, 2023-Palvannan, 2025-Tufano}]
        [Human-AI Alignment (5)\\ \cite{2022-Madi, 2025-Alami, 2025-Alami-TOSEM, 2025-Salem, 2024-Watanabe}]
      ]
    ]
  ]
\end{forest}
\end{adjustbox}
\caption{Taxonomy tree of MCR research topics and literature.}
\label{fig:taxonomy}
\end{figure*}

To ensure consistency with our research taxonomy defined in Section \ref{Sec-2}, we categorize the primary contributions of the 327 surveyed papers into two overarching directions: \textit{Improvement Techniques} and \textit{Understanding Studies}. Table \ref{Table 3} provides the definitions for these categories and the distribution of the surveyed literature. To further illustrate the granularity of these research efforts, Figure \ref{fig:taxonomy} presents a comprehensive taxonomy tree, mapping the major directions to specific sub-categories and their associated literature. The corresponding categorizations, associated papers, and reference links have been systematically curated and are publicly available in our GitHub repository\footnote{\url{https://github.com/watreyoung/MCR-Survey}}.

As shown in Table \ref{Table 3}, the research effort is remarkably balanced between the two directions. Specifically, \textbf{Improvement Techniques} account for 153 papers (46.8\%), focusing on the development of automation tools and algorithmic optimizations to enhance review efficiency. As detailed in Figure \ref{fig:taxonomy}, this direction is driven by five core areas, with ``Code Change Analysis'' and ``Unified Automation Frameworks'' emerging as the most populated sub-categories, reflecting the community's recent pivot toward deep-learning and LLM-based solutions.
On the other hand, \textbf{Understanding Studies} comprise 174 papers (53.2\%), spanning from large-scale repository mining to in-depth psychological and social investigations. The taxonomy reveals a high concentration of research in ``Process Efficiency'' and ``Information Needs,'' alongside a burgeoning interest in ``Human-AI Collaboration''—a sub-field that bridges empirical understanding with next-generation automation.

This equilibrium suggests that the MCR community places equal importance on the dual goals of technological innovation and empirical self-reflection. The balanced distribution further validates our roadmap's structure, which seeks to bridge the gap between understanding the ``human-centric process'' and building the ``AI-driven future.'' By organizing the literature into this hierarchical structure, we provide a clear navigation map for researchers to identify well-explored domains versus emerging gaps in the current MCR landscape.

\section{Current Status of MCR Research.}
\label{Sec-4}

\subsection{Improvement Techniques}
\subsubsection{Code Change Analysis.}
Code change analysis techniques aim to assist both contributors and reviewers in obtaining in-depth insight into code changes more comprehensively and more efficiently. Specifically, this technology can be divided into six subcategories: change decomposition, change understanding, specific change identification, defective change prediction, change prioritization, and change improvement.

\textbf{Change Decomposition.} Change decomposition aims to partition a complex or big change, which usually consists of multiple targets (e.g., bug fixes, feature additions, and refactoring) \cite{2015-Tao}, into several simpler and smaller ones (i.e., atomic changes). These approaches range from utilizing def-use relationships and heuristic patterns to identify cohesive change clusters \cite{2015-Barnett-ICSE,2015-Tao,2018-Freire}, to applying program slicing and comprehensive dependency analysis that considers inheritance and method-override relationships for more accurate partitioning \cite{2019-Guo,2019-Wang-ASE}. Such decomposition techniques have been shown to facilitate reviewer understanding while reducing false positive defects and increasing improvement suggestions \cite{2019-Biase}.

\textbf{Change Understanding.} To help developers and reviewers better comprehend the implications of code changes, some analytical approaches have been developed. These include \textbf{design-level techniques} for identifying design inconsistencies and predicting design-impactful modifications \cite{2013-He,2021-Uchoa-MSR}, alongside \textbf{impact analysis} that traces semantic relationships via AST-based methods \cite{2018-Wen,2019-Hanam} or summarizes \textbf{behavioral differences} to highlight effect changes \cite{2017-Menarini}. Complementary efforts employ commit clustering \cite{2014-Yamauchi} and patch linkage detection \cite{2021-Wang-IST} to clarify implementation intents. \yzz{Addressing cognitive challenges beyond textual differences, recent research advances visual interfaces, ranging from enhanced diffing of refactorings \cite{2024-Heander} and execution traces \cite{2022-Kanda}, to diagrammatic visualizations of structural dependencies and risks \cite{2017-Wang, 2021-Balci, 2023-Fregnan}, and finally software city metaphors \cite{2015-Tymchuk, 2024-KrauseGlau}.} Furthermore, contextualized environments are proposed to synthesize diverse artifacts, effectively bridging the gap between low-level diffs and high-level decision-making \cite{2024-Unterkalmsteiner}.

\textbf{Specific Change Identification.} Identifying the specific type of change according to different review requirements can help the reviewer adopt appropriate actions to improve the review efficiency. Approaches for identifying \textbf{recurring or systematic changes} leverage generalized edit patterns \cite{2013-A-Meng,2015-Zhang}, \textbf{clone evolution visualization \cite{2018-Fish}}, and deep learning to detect inconsistencies and anomalies \cite{2020-Ayinala}. For complex multi-file commits, learning-based methods extract \textbf{salient classes} to reveal the core essence of modifications \cite{2022-Huang}. Recent research targets granular attributes to optimize review management, such as assessing \textbf{code change reviewability} \cite{2018-Ram}, predicting \textbf{Large-Review-Effort (LRE) changes} \cite{2019-Wang}, and forecasting \textbf{revision necessity} \cite{2019-Shi} to guide resource allocation. To ensure integrity, recent models flag \textbf{high-complexity changes} \cite{2024-Willenbring}, recommend \textbf{missing co-changed functions} via Graph Neural Networks (GNN) \cite{2024-Hong}, and enforce \textbf{policy compliance} by verifying guarantees \cite{2023-Afzali}, auditing \textbf{dependency review coverage} \cite{2023-Imtiaz}, or automating \textbf{compliance assessment} via multi-task pre-training \cite{2023-Sawant}. Additionally, specialized tools streamline code evolution assessment by detecting \textbf{incomplete refactoring} \cite{2017-Chen}, distinguishing \textbf{refactoring parts} via replay and validation \cite{2017-Ge}, and reducing cognitive effort through intelligent diffs \cite{2021-Brito}, while recent work classifies \textbf{refactoring reviews} based on multi-dimensional criteria \cite{sieber2025classification}.

\textbf{Defective Change Prediction.} This task aims to identify whether a change is buggy or clean, providing an extra perspective of review and serving as an indicator to guide the allocation of effort of reviewers. Early feature-based approaches utilize traditional classifiers on process and developer metrics \cite{2016-Soltanifar,2017-Lal,2017-Madera}, later augmented by external knowledge from Stack Overflow \cite{2019-Sodhi,2021-Kapur}. Recent advancements integrate formal specifications \cite{2021-Liu} and multi-step localization frameworks \cite{2023-Sghaier}. \yzz{Driving towards finer granularity, interpretable models now recommend specific defective lines \cite{2022-Hong-SANER} or pinpoint problematic tokens via CodeBERT \cite{2024-Morikawa}. Specialized solutions have also extended to non-textual Low-Code platforms \cite{2024-choudhury-investigation} and verification artifacts, such as predicting test plan quality to prevent reversions in industrial environments \cite{2022-Chen}.} Furthermore, LLM-driven frameworks integrate automated penetration testing to detect prompt injections \cite{takieldeen2025unlocking}, while approaches like iCodeReviewer \cite{2025-Peng} employ a Mixture-of-Prompts strategy to mitigate hallucinations and enhance vulnerability localization.

\textbf{Change Prioritization.} Change prioritization addresses a critical challenge of determining the optimal order for reviewing incoming changes to maximize reviewer efficiency \cite{2015-O-Gousios}. The approaches have evolved from early bug-proneness optimization \cite{2013-Aman} to machine learning prioritizers integrating contributor tracks and social factors \cite{2015-A-Veen,2019-A-Zhao}, and predictive models for merge likelihood to avoid wasted effort \cite{2018-Fan}. Recent advancements employ expert-guided feature identification \cite{2021-Saini}, algorithms like LightGBM \cite{2022-Islam}, and cost-aware multi-objective frameworks \cite{2022-Huang-ESE,2024-Chouchen}. \yzz{The state-of-the-art techniques further utilize LLM embeddings to predict review ``hot-spots'' \cite{2024-Olewicki} and apply GNN to model complex interdependencies within multigraph structures \cite{2024-Yang}. Finally, addressing the limitations of static metrics, discrete-event simulations have been proposed to model the dynamic review process, validating that Random Forest prioritizers significantly outperform rule-based baselines in delivery efficiency \cite{2025-Yang}.}

\textbf{Change Improvement.} Change improvement technologies aim to detect and fix issues in patches before submission to reviewers, allowing reviewers to focus on more complex problems. These methodologies have transitioned from early AST-based auto-correction tools like \textit{Fix-it} \cite{2013-Balachandran} and pattern-based convention mining \cite{2019-Ueda} to deep learning generation. The existing initial vocabulary constraints \cite{2021-Tufano} are mitigated via Byte-Pair Encoding \cite{2022-Thongtanunam}, followed by the integration of structural awareness \cite{2023-Lin} and diff-aware mechanisms \cite{2023-Pornprasit} to enhance precision. \yzz{Addressing single-phase limitations, CodeReviser \cite{2025-Wang} employs a ``Divide-and-Conquer'' strategy that separates modification localization from revision execution. Industrial viability is evidenced by Google's large-scale deployment of a T5-based patch generator \cite{2024-FrommgenA24}. Furthermore, general-purpose LLMs have demonstrated the capability to outperform specialized baselines through context-rich zero-shot prompting \cite{2024-Guo}.}

\subsubsection{Reviewer Recommendation.} Reviewer recommendation aims to identify the most suitable experts for a given code change. Existing approaches can be broadly categorized into three streams based on the primary factors they utilize: technical factors, non-technical factors, and hybrid approaches.

\textbf{Technical Factors.} These approaches primarily focus on the alignment between the technical content of the code change and the historical expertise of the reviewers. 
\begin{itemize}
    \item \textit{Change characteristics.} The methodologies have progressed from matching historical line modifications \cite{2013-Balachandran-ICSE} and file paths \cite{2015-Thongtanunam-SANER} to integrating multi-dimensional features. Specifically, the approaches combine text mining with path analysis to capture semantic and structural similarities \cite{2015-Xia}, with recent findings confirming that fusing path and semantic similarity maximizes recommendation accuracy \cite{2023-Li}.

    \item \textit{Reviewers' experience.} Expertise modeling \cite{2016-Hannebauer} has expanded from search-based optimization of collaboration patterns \cite{2016-Ouni} and cross-project experience \cite{2016-Rahman-ICSE} to nuanced metrics like comment contribution \cite{2016-Zanjani} and patch-specific effort quantified through size and complexity variables \cite{2014-Mishra}. Subsequent frameworks integrate diverse factors including responsiveness and code context \cite{2020-Sadman}. \yzz{Recently, focus has shifted to data reliability: techniques now mitigate labeling noise from availability bias \cite{2021-Tecimer} and prioritize substantive feedback by filtering for Quality-Enhancing Review Comments \cite{2024-Rong}.}
\end{itemize}

\textbf{Non-Technical Factors.} These approaches address the inherent collaborative nature of MCR by considering social dynamics and human resource constraints.
\begin{itemize}
    \item \textit{Social relations.} Social relation-based approaches leverage collaborative patterns among reviewers to enhance recommendation accuracy. Early research highlights the efficacy of comment networks \cite{2016-Yu} and engagement metrics \cite{2017-Jiang} over complex attributes. Recent advancements apply graph representation learning to socio-technical networks, utilizing GNN to identify qualified reviewers with implicit expertise, thereby overcoming the limitations of history-based interaction matching \cite{2023-Zhang}.
    \item \textit{Workload balancing.} These methods integrate load distribution mechanisms to mitigate unbalanced assignments and reduce review latency \cite{2019-Asthana}. Multi-objective frameworks optimize trade-offs between expertise, availability, and collaboration \cite{2020-Rebai}, though utility can be limited by developers' predetermined choices \cite{2020-Kovalenko}. Complementary efforts include network visualizations for turnover management via expertise preservation \cite{2024-Hajari}. \yzz{Industrially, Meta's large-scale A/B tests demonstrate that workload-aware re-ranking effectively reduces burden, while explicit random assignment significantly decreases latency by mitigating the bystander effect \cite{2025-Rigby}.}
\end{itemize}

\textbf{Hybrid Approaches.} Beyond traditional feature-based methods, integrated approaches leverage complex modeling techniques to capture comprehensive reviewer-change relationships. Hybrid strategies combine multiple recommendation components, from integrating latent factor models with neighborhood models to uncover implicit relations \cite{2017-Xia}, to multi-module architectures that orchestrate similarity matching, classification, and filtering \cite{2022-Pandya}. Deep learning pipelines employ masked language models for quality prediction and reviewer-commit distance measurement \cite{2022-Aryendu}, with industrial evaluations confirming the importance of collaboration and cross-project experience \cite{2022-Kong}. In addition, graph-based representations model the inherently relational nature of code review, utilizing hypergraphs to capture high-order multi-reviewer interactions \cite{2022-Rong} \yzz{and multiplex connectivity across diverse development roles (e.g., committers, commenters) \cite{2024-Qiao}}, integrating code semantics extracted by Transformer models into graph embeddings through attentive propagation \cite{2023-Liu}, and constructing traceability graphs that connect diverse software artifacts for comprehensive reviewer familiarity assessment \cite{2019-Suluun,2021-Sulun}.

\subsubsection{Review Comment Synthesis.} It requires extensive programming and review experience to write high-quality review comments. Review comment synthesis techniques facilitate reviewers by mining and recommending useful comments from code evolution history. These approaches have evolved through distinct paradigms.

\textbf{Retrieval-Based Methods} focus on matching current changes with historical data. While early techniques mine association rules \cite{2018-Chatley}, subsequent studies reveal that simple Bag-of-Words retrieval outperforms complex deep learning models, challenging assumptions about necessary complexity \cite{2022-Hong-FSE}. To address granularity limitations, approaches incorporate structured elements (e.g., file paths) to support broader recommendation scopes \cite{2023-Shuvo}. \yzz{Recently, pre-trained transformer-based models, such as UniXcoder, have advanced this paradigm by optimizing vectorization efficiency \cite{2024-Kartal}, while platforms like ACoRA leverage advanced Transformers (e.g., CodeT5+) to retrieve similar fragments for automated, taxonomy-based review guidance \cite{ochodek2025acora}.}


\textbf{Generation-Based Methods} synthesize novel feedback directly from code changes by learning the semantic mapping between code and comments. This paradigm has evolved from specialized neural architectures to general-purpose LLMs:

    \begin{itemize} 
        \item \textit{Deep learning-based methods.} These approaches employ neural architectures to map code changes to reviews. Early models utilize multi-encoder LSTMs for contextual understanding \cite{2018-Gupta}, evolving into bidirectional architectures with attention mechanisms to capture deeper semantics \cite{2020-Siow}. Additionally, hybrid strategies combine classification with retrieval, identifying necessary explanations via TF-IDF before retrieving analogous examples \cite{2022-Rahman}.
    
        \item \textit{LLM-based methods.} \yzz{Related research highlights distinct optimization requirements: structural representations benefit fine-tuned models, whereas frozen LLMs rely on inherent contextual inference \cite{2024-Lu}. To address contextual limitations, hybrid frameworks synergize LLMs with static analysis via Retrieval-Augmented Generation (RAG) \cite{2025-Jaoua} or integrate AST-based expansion with Chain-of-Thought guidance \cite{2025-Zhang}. For finer granularity, CodeDoctor employs a gated fusion decoder to synthesize multi-perspective feedback from category-specific pairs \cite{2025-Li}. Prioritizing data quality, recent studies mitigate noise by integrating reviewer expertise into loss functions \cite{2025-Lin} or applying LLM-based semantic filtering, validating that high-fidelity data yields superior performance \cite{2025-Liu}.}
    \end{itemize}

\subsubsection{Review Comment Analysis.}

Review comment analysis techniques evaluate the quality and content of review comments to enhance review efficiency and maintain healthy collaboration. These approaches primarily address three dimensions: analyzing content and intent, detecting sentiment patterns, and assessing quality.

\textbf{Content and Intent Analysis.} This dimension mines unstructured text to comprehend the semantic subject and purpose of feedback. At the entity level, hybrid named entity recognition (NER) frameworks integrate regex pre-tokenization with pre-trained models to robustly extract technical elements despite syntactic noise \cite{2024-Kachanov}. At the semantic level, methodologies evolve from BERT-based taxonomy classification \cite{2022-Ochodek} to macroscopic topic modeling. Notably, recent innovations leverage LLMs to automate topic labeling, effectively bridging the gap between embedding-based clustering performance and human interpretability \cite{2025-Iftikhar}. Addressing data sparsity and fragmentation, Petrova et al. \cite{petrova2025building} unify existing datasets into a standardized hierarchical scheme (``CleanCodeReview''), demonstrating that domain-specific transformers like CodeBERT significantly outperform general embeddings for classifying diverse review intents.

\textbf{Sentiment and Tone Analysis.} Analyzing emotional polarity mitigates interpersonal conflict. The methodologies have progressed from TF-IDF baselines \cite{2017-Ahmed} to the precise detection of negative dynamics. \yzz{While Egelman et al. \cite{2020-Egelman} leverage process metrics (e.g., shepherding time) to predict ``pushback'' with high recall, the limited precision of such methods prompts a shift toward text analysis for identifying toxic spans \cite{2023-Sarker} and linguistic pushback features \cite{2022-Qiu}.} Transitioning from detection to active intervention, recent frameworks fine-tune Transformers (e.g., T5) on ChatGPT-augmented data to automatically neutralize toxicity into civil alternatives while preserving semantics \cite{2024-Rahman}. Beyond textual modalities, physiological studies analyze nonverbal cues (e.g., eye gaze) to reveal genuine in-situ sentiment \cite{2020-Vrzakova}.

\textbf{Quality and Usefulness Assessment.} This task evaluates the constructive value of feedback. Early methods utilize vector-based semantic similarity \cite{2014-Pangsakulyanont}. Machine learning approaches advance this by leveraging multi-dimensional features to predict utility \cite{2017-Rahman,2021-Hasan} and detect specific communication hurdles such as reviewer confusion \cite{2017-Ebert}. Recent innovations apply multi-label learning for comprehensive attribute analysis \cite{2023-Yang}. Reflecting on this evolution, a comprehensive survey highlights the progression from heuristic definitions—such as code change triggers \cite{2015-A-Bosu} and semantic similarity \cite{2014-Pangsakulyanont}—to robust prediction models that, despite data availability constraints, increasingly rely on sophisticated feature engineering and linguistic semantics to assess utility \cite{2023-Ahmed}. Furthermore, fine-tuned LLMs (e.g., GPT-4o) achieve state-of-the-art performance, outperforming surprisingly effective Bag-of-Words baselines, though domain generalizability remains a challenge \cite{2025-Ahmed}. 
\subsubsection{Unified Automation Frameworks.}

Unlike task-specific approaches, unified automation frameworks leverage shared representations and architectures to address multiple code review activities within a single system. These frameworks reduce the fragmentation of review tools and enable more seamless automation across the entire review workflow.

\textbf{Deep Learning-Based Methods.} The development of unified frameworks begins with adapting general-purpose transformers for review tasks. T5-based models demonstrate initial success in jointly generating improved code and review comments from raw source code, eliminating the need for code abstraction \cite{2022-Tufano}. This capability advances significantly with CodeReviewer \cite{2022-Li}, a CodeT5-based model pre-trained specifically on large-scale review data, which simultaneously handles code quality estimation, comment generation, and code refinement tasks with superior performance. To address the complexity of code review which involves interrelated activities, researchers have designed architectures that optimize multiple objectives simultaneously. Structure-aware methods like PDG2Seq capture program dependencies beyond sequential code representation, with crBERT successfully fusing semantic and structural features to improve automation performance across multiple review tasks \cite{2023-Yin}. Multi-perspective approaches, such as SMILER, develop dual models tailored for different stakeholders, with developer models generating code improvement while reviewer models incorporate comments \cite{2024-Cao}. In addition, MULTICR predicts review outcomes by simultaneously optimizing for both merged and abandoned change recall, producing interpretable IF-THEN rules that balance performance with transparency \cite{2024-Chouchen-TOSEM}. \yzz{AutoCommenter \cite{2024-Vijayvergiya} leverages a multi-task T5 model to automate best practice assessment, highlighting that practical success requires not only vast training corpora but also robust inference strategies, such as dynamic thresholding and beam search, to mitigate false positives and ensure user acceptance. In parallel, to enhance accessibility for novice developers, recent tools integrate CodeBERT for error classification and Refined CodeT5 for repair within intuitive GUIs, providing structured, two-step feedback \cite{2024-Xuan}. To further exploit the dependencies between successive tasks, DISCOREV leverages cross-task knowledge distillation, utilizing a student-teacher architecture where downstream assessments explicitly guide the optimization of upstream generation models \cite{2024-Sghaier}. While the aforementioned studies focus on model design, the CuRev pipeline highlights that data quality is equally critical for multi-task success. By utilizing an LLM-as-a-Judge to filter and reformulate noisy reviews, it constructs a high-quality corpus that enables fine-tuned LLMs to achieve significant performance gains in both comment generation and code refinement tasks \cite{2025-Sghaier}.}

\textbf{LLM-Based Methods.} \yzz{Most recently, the research focus has shifted toward the accessibility and interactivity of LLMs. On the one hand, parameter-efficient fine-tuning (PEFT) techniques demonstrate that effective review automation need not require massive computational resources \cite{2023-Lu,2024-Yu-TOSEM}. On the other hand, the generative capabilities of LLMs facilitate interactive workflows that enhance developer engagement, ranging from conversational assistants \cite{2024-Almeida, 2024-chowdhury} and gamified frameworks \cite{2024-Khelifi}, to autonomous multi-agent systems that simulate role-based team collaboration for comprehensive review tasks while utilizing supervisory mechanisms to prevent prompt drifting \cite{2024-Tang}. Further advancing this multi-agent paradigm, HYDRA-REVIEWER \cite{ren2025hydra} orchestrates specialized agents based on a fine-grained taxonomy to address the lack of multi-dimensional analysis, employing context retrieval and reflective mechanisms to mitigate vagueness and hallucinations while prioritizing critical issues for developers. In the realm of green software engineering, an adaptive multi-agent system utilizes reinforcement learning to integrate explainable, real-time energy optimization into workflows, balancing energy savings with developer acceptance \cite{sharanarthi2025adaptive}. From an industrial perspective, Ericsson reports on a lightweight industrial implementation that integrates open-source LLMs (e.g., Code Llama) with static analysis for context extraction, demonstrating that effective prompt engineering can serve as a cost-efficient alternative to expensive fine-tuning for generating relevant reviews in enterprise environments \cite{2025-Ramesh}. The BitsAI-CR framework \cite{2025-Sun} of ByteDance tackles the precision and practicality challenges of large-scale deployment by implementing a taxonomy-guided two-stage pipeline (generation and filtering) alongside a data flywheel mechanism that leverages a novel ``Outdated Rate'' metric for continuous optimization. Moreover, to enhance adaptability, recent frameworks integrate RAG to establish long-term memory, enabling agents to provide personalized and self-refining feedback \cite{sharanarthi2025multi}.}

\begin{center}
    \begin{tcolorbox}[colback=gray!10!white,colframe=black!75!black,title=\textbf{Summary for Improvement Techniques}]
            \begin{enumerate}
    
                \item \textbf{Code change analysis} techniques assist in decomposing, understanding, and prioritizing modifications. Recent advancements utilize structure-aware models and LLMs to extend these capabilities into security vulnerability detection and automated code refinement.

                \item \textbf{Reviewer recommendation} approaches have evolved from solely matching technical expertise to integrating non-technical factors, such as social relations and workload balancing, to optimize collaboration dynamics and reduce review latency.                

                \item \textbf{Review comment processing} has shifted from traditional retrieval and specialized deep learning models towards using RAG and generative LLMs for synthesizing constructive feedback. Concurrently, analysis techniques are increasingly focused on assessing comment usefulness and actively mitigating toxic sentiments to foster healthy communication.                

                \item \textbf{Unified automation frameworks} represent the latest paradigm, addressing multiple review tasks within a single system. The field is transitioning from multi-objective pre-trained models to autonomous multi-agent systems and scalable industrial solutions that streamline the entire workflow.
            \end{enumerate}
    \end{tcolorbox}
\end{center}

\subsection{Understanding Studies}

\subsubsection{Quality Assurance and Reliability}
This section examines the effectiveness of code review in ensuring software quality, addressing defect detection capabilities, security assurance, and the handling of specialized artifacts.

\textbf{Defect and Risk Detection.} Defect detection remains a primary motivation for code review. Research employs review intensity and participation levels as measurement criteria \cite{2024-Yu-ESE}, uncovering that historically risky and future-defective files often receive less rigorous review than clean counterparts \cite{2015-Thongtanunam,2017-Izquierdo-Cortazar17}.
Automated code analysis tools complement human efforts with varying effectiveness. While static analysis tools like PMD reduce review effort by enforcing warning removal, they generate false positives \cite{2015-Panichella} and address only approximately 16\% of issues identified in manual reviews \cite{2017-Singh}. Expanding on this, Gunawardena et al. \cite{2023-Gunawardena} categorize 116 specific defect types, suggesting that pre-review automated detection can significantly reduce downstream review costs. To bridge this gap, project-specific customization of rule sets is essential to detect context-sensitive violations \cite{2023-Mehrpour}. Furthermore, formal verification tools significantly increase review effectiveness, suggesting broader adoption beyond safety-critical systems \cite{2016-Hentschel}. \yzz{From an educational perspective, coding style checkers prevent the recurrence of automatically detected issues, whereas manually identified problems tend to persist \cite{2018-Ueda}.}

\textbf{Security Review Effectiveness.} Despite the priority placed on defect finding, security considerations reveal significant gaps. Yu et al. \cite{2023-Yu} show that security issues are not frequently discussed in the process of code review. Only approximately 1\% of review comments address security issues \cite{2013-Bacchelli,2016-O-Biase}. While specialized knowledge enables the detection of complex issues like race conditions \cite{2023-Alfadel}, many identified concerns are merely acknowledged without fixes or merged despite unresolved discussions \cite{2024-Charoenwet-ESE}. These findings suggest increasing reviewer numbers beyond the typical two-reviewer model for security-critical changes \cite{2023-Alfadel}. \yzz{Recent evaluations show that while security-focused static analysis tools effectively prioritize review efforts, their utility is constrained by high false positive rates and blind spots for certain vulnerability patterns \cite{2024-Charoenwet-ISSTA}. Deep learning systems have demonstrated strong adaptability to these security-specific scenarios \cite{2018-Wang}, and interactive approaches like ``serious games'' actively bridge knowledge gaps \cite{2024-AndreiCristian}.} From the practitioner's perspective, security knowledge gaps are critical, as developers unintentionally deprioritize security issues despite explicit organizational requirements \cite{2022-Braz-FSE}.

\textbf{Reviewing Specialized Artifacts.} Different code artifacts require distinct quality standards and review approaches.
\begin{itemize}
    \item \textit{Documentation and Specs.} Build specifications receive disproportionately less attention despite correctness implications \cite{2023-Nejati, 2024-Rokem}. For documentation, comments create unique dynamics that aid understanding but may trigger additional updates \cite{2022-Rao}.
    \item \textit{Test Code:} Test files exhibit defect patterns distinct from production code, yet garner less discussion due to limited navigation support. \cite{2018-Spadini}. Therefore, Test-Driven Review (TDR) is proposed to increase bug discovery in tests \cite{2019-Spadini}.
    \item \textit{Refactoring Challenges.} Refactoring changes are inherently risky, exhibiting higher crash-proneness \cite{2018-An} and requiring significantly more review effort \cite{2022-AlOmar}. This burden is often exacerbated by inadequate documentation of refactoring intent, which creates comprehension barriers when structural changes are mixed with feature additions \cite{2020-Paixao,2021-AlOmar}.
    \item \textit{Semantics and Maintainability.} While developers often rely on static analysis to address general code smells \cite{2020-Han,2022-Han,2024-Tuna}, automated tools provide limited coverage for ``understandability smells'' \cite{2025-Oliveira}. Similarly, ensuring method naming consistency requires semantic nuances that tools often miss \cite{2025-Kim}, highlighting the persistent need for human judgment in maintaining code readability.
    \item \textit{Strategies for Structural Quality.} To optimize the review of structural changes, qualitative analysis reveals that senior reviewers play a pivotal role, actively inducing nearly 45\% of validated refactorings by guiding authors on structural logic \cite{2025-Coelho}. Furthermore, utilizing a dedicated ``Refactor'' branch allows reviewers to focus specifically on internal quality attributes, effectively mitigating review costs \cite{2025-AlOmar}.
    \item \textit{Emerging Domains.} GUI-based testing artifacts benefit from adapted guidelines bridging traditional review and interface testing \cite{2023-Bauser}. Similarly, research software teams—often constrained by informal processes and limited personnel—are urged to adopt formal peer review to ensure code maintainability and reproducibility \cite{2022-Eisty}. \yzz{Infrastructure-as-Code (IaC) necessitates tailored criteria prioritizing logic and dependencies over standard metrics \cite{2025-Bessghaier}. In safety-critical domains like aerospace, integrating automated reviews reduces technical debt, though workflows must balance continuous pipelines with periodic gates to maintain compliance \cite{ludwig2025enhancing}.}
\end{itemize}

\subsubsection{Process Efficiency and Workflow Patterns}
Efficiency is a critical dimension of MCR, encompassing development velocity, strategic choices, and the management of process bottlenecks. Notably, a misalignment persists: while researchers often focus on human factors, practitioners prioritize review process properties and quality impacts \cite{2023-Badampudi}.

\textbf{Code Velocity and Turnaround Time.} Code velocity measures the rate at which changes move through the review process. Time-to-merge is identified as a critical metric \cite{2024-Kudrjavets}, with patch size and author experience serving as key factors—smaller changes and experienced authors consistently achieve faster turnaround \cite{2022-Kudrjavets, 2023-Kudrjavets}. Distributed development contexts present trade-offs where multi-team participation improves quality but significantly extends review duration \cite{2018-Santos}; such asynchronicity frequently necessitates rebasing, which triggers developer rework and complicates empirical data analysis \cite{2019-Paixao}. Crucially, review response time is further modulated by organizational affiliation and reviewer activity, with statistical evidence from WebKit and Blink confirming that intra-organizational reviews (e.g., Apple-to-Apple) and highly active reviewers yield significantly faster feedback \cite{2016-Baysal}. Early-week submissions effectively mitigate weekend delays \cite{2022-Chen-FSE}, with predictive models identifying reviewer activity and interaction history as primary determinants of review duration \cite{2022-Jiang,2023-Chouchen-ESE-L}, alongside repository features like contributor count that underscore the importance of workflow regularity \cite{2022-Al-Rubaye}. \yzz{Recently, Kansab et al. \cite{2025-Kansab} demonstrate that excluding anomalies like ``Merge Request deviations'' is essential for unbiased completion time prediction. Underpinning these empirical evaluations are robust data infrastructures, exemplified by industrial real-time monitoring platforms like \textit{CodeFlow Analytics} \cite{2015-Bird} and open-source mining tools like \textit{GERRITMINER} \cite{2018-Spadini}.}

\textbf{Review Strategies and Change Taxonomy.} No review strategy exhibits universal superiority. Broadcast reviews achieve twice the speed of unicast reviews but sacrifice quality \cite{2017-Armstrong}. Pre-commit reviews optimize efficiency metrics compared to the quality-focused post-commit approach \cite{2016-Baum-ICSSP}. Sequential reviews foster learning, whereas parallel reviews increase throughput \cite{2018-Luxton-Reilly}. \yzz{In education, pair review is preferred for interactivity, while individual review proves superior for skill acquisition \cite{2025-Koitz-Hristov}.} In the process of code review, checklist effectiveness varies, benefiting complex reviews but potentially reducing performance in simpler tasks \cite{2022-Braz,2022-Goncalves-ESE}. Furthermore, cognitive biases regarding information presentation are evident, as the presentation order of files significantly impacts comment density, with files displayed earlier receiving disproportionate attention \cite{2022-Fregnan-FSE-order}. To enable targeted strategies that align with specific contexts, researchers focus on classifying change types. \yzz{Empirical studies reveal that the majority of review-triggered changes address system evolvability rather than functionality \cite{2014-Beller, 2025-Davila, 2022-Fregnan-ESE-evoluation}, with initial patch size significantly influencing the volume of subsequent modifications \cite{2022-Fregnan-ESE-evoluation}.} Surveys from the developer's perspective further refine these taxonomies to identify specific automation needs across different change types \cite{2020-Panichella}. Industrial schemes encompass 20 distinct facets for context-specific selection \cite{2016-Baum-QRS}, and deep learning approaches now automate this classification for intelligent routing \cite{2022-Fregnan-ESE-classify,2023-Turzo}. Collectively, these findings suggest that no single review strategy exhibits universal superiority.

\textbf{Information Needs and Supporting Materials.} Throughout the review workflow, reviewers exhibit distinct information needs, ranging from alternative solution evaluation to verification \cite{2018-Ebert,2018-Pascarella}. \yzz{Cognitive processes underpin the interpretation of such information. Specifically, working memory capacity affects the detection of delocalized defects \cite{2019-Baum}, and reviewers employ opportunistic strategies like dynamic scoping to construct mental models \cite{2025-Goncalves}.} Studies consistently demonstrate that additional documentation, such as UML diagrams and software metrics, enhances effectiveness \cite{2019-Belgamo, 2015-Rong}, while review linkages facilitate context understanding \cite{2019-Hirao-FSE,2021-Wang-ESE}. \yzz{Regarding feedback forms, code snippets are predominantly utilized for implementation suggestions \cite{2024-Zhang,2019-Hirao,2022-Fu}.} However, design-related discussions are surprisingly rare \cite{2017-Li,2018-Zanaty}. Reviewer confusion is often exacerbated by systematic information failures, including missing rationale \cite{2021-Ebert}, and nearly 29\% of comments lack sufficient relevance \cite{2025-Chen}, with actionable feedback quality further contingent on factors like reviewer selection and changeset management \cite{2015-A-Bosu}. To overcome adoption barriers, studies suggest that integrating reviews into existing submission workflows and minimizing context switching are critical for industrial acceptance \cite{2017-Baum-PROFES, 2017-Baum}, echoing developer perspectives that cite tool support, time management, and context switching as primary challenges impacting review quality \cite{2016-Kononenko}. Frameworks have been proposed to evaluate comment effectiveness through linguistic characteristics like politeness and comprehensibility \cite{2024-Turzo}. Recently, LLMs have shown promise in automatically categorizing these interactions \cite{2024-Martin-ASE} and generating rationale explanations \cite{2025-Widyasari}.

\textbf{Process Bottlenecks and Failures.} Understanding the root causes of failure is paramount for optimizing review workflows. Specifically, poor participation implies a bottleneck often driven by sparse descriptions and limited interaction history \cite{2017-Thongtanunam}, while abandonment patterns primarily stem from duplicate changes, and rejected contributions correlate with incivility \cite{2019-Wang-IST,2021-Ferreira}. \yzz{To mitigate such delays, nudging mechanisms like \textit{NudgeBot} \cite{2022-Shan} successfully reduce review time by prioritizing notifications for reviewers most likely to act based on historical participation.}
Complex modifications, such as tangled code changes, frequently obscure vulnerabilities \cite{2021-Paul-ICSE}. \yzz{Superficial approvals, or ``LGTM smells'', are prone to quality issues \cite{2024-Gon}, though recent tools integrating mutation testing have proven effective in mitigating these shallow reviews \cite{2023-Mukhtarov}.} Reviewer cognitive limitations, like perceptual confusion, lead to predictable defect detection failures \cite{2024-Huang-JSS}. Beyond communication, misalignments between review tools and participants' perspectives further degrade the developer experience \cite{2022-Soderberg}.

\subsubsection{Human Factors and Social Interaction}
MCR is inherently a socio-technical process. This section explores how human cognition, social interactions, and demographic factors shape review outcomes.

\textbf{Demographic and Social Influence.} Developer reputation and organizational affiliation create disparities in review experiences \cite{2013-Baysal,2014-Bosu}, yet not all demographic traits are predictive; for instance, reviewer age does not correlate with efficiency or correctness despite its association with experience \cite{2017-Murakami}. Gender bias emerges as a systemic issue, where women face reduced opportunities for reviewer selection \cite{2023-Murphy-Hill,2023-Sultana} and receive disproportionately critical feedback \cite{2019-Paul}. \yzz{Recent evidence indicates that gender-linked cognitive style mismatches result in lower feature discoverability for women, mitigable through inclusive UI design \cite{2024-Murphy-Hill}.} Social interaction patterns, such as collaborator familiarity and participation levels, further influence review quality \cite{2014-Meneely,2014-Rigby,2015-Kononenko,2021-Wang}, with metrics like historical review participation rate proving critical for predicting future reviewer engagement \cite{2019-Ruangwan}. Addressing these biases, anonymous author code review is explored as a mitigation strategy, though it introduces communication barriers due to information suppression \cite{2022-Murphy-Hill-TSE}. \yzz{In decentralized contexts, blockchain-based incentives aim to boost participation but risk deterring contributors due to high operational costs \cite{martins2026tokenized}.} More broadly, practitioners warn that such systemic lack of fairness in the review process—real or perceived—can demoralize contributors and threaten long-term project sustainability \cite{2018-German}. Cultural and social issues often outweigh technical concerns as obstacles to code review adoption \cite{2016-Baum}, requiring careful balancing of competing priorities within organizations \cite{2017-Macleod}.

\textbf{Biometrics and Cognitive Factors.} Physiological measurements provide objective insights into reviewer behavior. Eye-tracking studies reveal that expert reviewers demonstrate superior error detection with reduced visual effort compared to novices \cite{2023-Hauser}. Providing a theoretical basis for these observations, a unified holistic model of visual perception adapts radiology and psychology frameworks (e.g., Global-Focal Search) to explain how experts utilize rapid ``global'' scans followed by ``focal'' analysis to optimize cognitive load during code review \cite{2025-Hauser}. \yzz{However, non-technical signals like urgency and reputation significantly alter visual attention distribution, highlighting a disconnect between self-reported factors and observed behaviors \cite{2024-Yabesi}. In platform interactions, developers prioritize core artifact information over meta-information but heavily rely on visual social signals like heatmaps \cite{2025-Wiese}.} Multi-modal biometric data (e.g., via \textit{iReview}) combining heart rate variability and pupillary response enables real-time quality assessment \cite{2021-Hijazi,2023-Hijazi}. Gaze-driven assistance platforms like GANDER leverage these insights, though initial deployments reveal user adaptation challenges \cite{2023-Saranpaa}. Cognitive biases, such as confirmation bias and decision fatigue, hinder objective reasoning \cite{2025-Jetzen}. \yzz{In education, novices struggle with ``delta comprehension'' due to tool limitations \cite{2024-Middleton,2025-Case}, necessitating AI-based scaffolding to mitigate cognitive load \cite{2025-Tseng,2025-Beattie,2025-Lee,2025-Duong}.}

\textbf{Communication and Collaboration.} The emotional aspects of code review significantly influence efficiency. Negative comments correlate with extended acceptance intervals \cite{2019-Asri}. Gender disparities manifest through asymmetric sentiment expression, where female developers express less negative sentiment yet receive reduced positive encouragement \cite{2019-Paul}. Furthermore, destructive criticism is prevalent, with non-specific negative feedback disproportionately impacting women and potentially contributing to the industry's gender diversity gap \cite{2022-Gunawardena}. Toxic interactions degrade both code quality and integration speed \cite{2021-Chouchen}, prompting the development of detection tools like \textit{ToxiCR} to automatically identify harmful comments \cite{2023-Sarker}. \yzz{Psychological research identifies that Code Review Anxiety (CRA) drives avoidance behaviors, though clinically-informed workshops can mitigate this distress \cite{2024-Lee}.} Conversely, visible comments and positive reminders enhance efficiency \cite{2020-Spadini}, a goal further supported by collaborative platforms like \textit{Fistbump} that streamline discussion coordination \cite{2016-Kalyan}. Reviewer disagreements present challenges but also opportunities; senior reviewers achieve higher agreement rates \cite{2016-Hirao}, and diverse viewpoints can be valuable despite efficiency costs \cite{2022-Hirao,2024-Goncalves}, as properly managed interpersonal conflicts often serve as catalysts for learning and improvement rather than mere obstacles \cite{2022-Goncalves}. Underpinning these dynamics, surveys highlight that while technical competencies are foundational, expert reviewers prioritize non-technical skills like clear communication and efficient collaboration to drive confidence and professionalism \cite{2023-Goncalves}. The emotional bonds within teams may compromise objective defect detection \cite{2023-Ciancarini}.

\textbf{Benefits of MCR.} Beyond defect detection, knowledge transfer is a primary benefit \cite{2013-Rigby,2017-Bosu,2021-Cunha}. Reviews facilitate expertise dissemination and implicit standardization of coding practices \cite{2021-Sri-Iesaranusorn,2022-Wen,2023-Iftikhar}, though empirical assessments have struggled to quantitatively capture the impact of this transfer on contribution quality, highlighting a need for further investigation \cite{2020-Caulo}. \yzz{Simulations model MCR as a time-varying hypergraph, showing that information can diffuse to over 85\% of participants in mid-sized projects within weeks \cite{2025-Dorner}.} Some organizations like Google prioritize readability and maintainability over strict defect detection \cite{2018-Sadowski}. Indeed, empirical studies confirm that stakeholders view engagement, team awareness, and communication as critical non-technical value drivers alongside the process's role in maintaining software stability \cite{2021-Cunha-SBQS}. Reviews also serve as architectural guardians \cite{2022-Li-ICSA} and trigger deeper quality discussions when combined with techniques like mutant testing \cite{2023-Petrovic}. \yzz{In education, gamified peer review and serious games validate this potential by enhancing student proficiency and self-confidence \cite{ismail2024systematic,2025-Ardic}.}

\subsubsection{Evolution toward Human-AI Collaboration}
The emergence of AI, particularly LLMs, is transforming MCR from a human-centric activity to a collaborative human-AI system. \yzz{Unlike distinct automation tools to help developers and reviewers, this paradigm shift positions AI as an active participant in the review workflow. To understand the implications of this collaboration, we examine the evolution through three critical lenses: AI capabilities and model performance, integration and practical challenges, and Human-AI alignment.}

\textbf{AI Capabilities and Model Performance.} Comparative evaluations establish the baseline capabilities of automated review. While early deep learning models like CodeT5 excel at specific tasks \cite{2023-Zhou}, recent LLMs demonstrate superior generalization \cite{2024-Guo}. \yzz{Parameter-Efficient Fine-Tuning (PEFT) effectively augments LLMs with code-change semantics, yielding superior performance over specialized small models \cite{2025-Fan}, addressing limitations in textual coherence observed in earlier bimodal architectures like CodeBERT \cite{gogri2025transformer}.} However, general-purpose LLMs can exhibit stochasticity. \yzz{SWR-Bench \cite{2025-Zeng} reveals that aggregating multiple independent reviews effectively mitigates instability.} Moreover, Tufano et al. \cite{2024-Tufano} highlight that existing models still struggle with complex code generation compared to humans, revealing critical limitations in refinement tasks. Specifically, fine-tuning yields the highest performance, while few-shot learning serves as a resource-efficient alternative \cite{2024-Pornprasit}.

\textbf{Integration and Practical Challenges.} Traditional review bots have successfully automated routine tasks, correlating with increased merge rates \cite{2020-Wessel,2022-Wessel}. However, the transition to generative AI introduces new challenges. While practitioners anticipate efficiency gains, adoption is tempered by concerns regarding trustworthiness and lack of project context \cite{2024-Davila}. Industrial case studies utilizing GPT-4 report that while tools enhance code quality, they paradoxically increase pull request closure duration due to the effort required to verify bot-generated comments \cite{2025-Cihan}. Controlled experiments confirm this, revealing that automated reviews do not save time and can bias reviewers toward lower-severity issues, as participants spend significant effort validating verbose LLM outputs without a corresponding increase in confidence or detection of critical defects \cite{2025-Tufano}. Conversely, other integrations of GPT packages have been shown to efficiently improve quality through timely feedback \cite{2023-Crandall}, highlighting the variability in implementation outcomes. Review bots also introduce challenges like communication noise and increased testing overhead \cite{2020-Wessel-SBES,2022-Kim}, though recent suggestion bots aim to minimize interruptions \cite{2023-Palvannan}.

\textbf{Human-AI Alignment.} The interaction between developers and AI is complex. \yzz{Eye-tracking reveals that AI-generated code receives reduced visual attention, indicating potential over-confidence that may compromise inspection quality \cite{2022-Madi}.} Furthermore, while LLMs mitigate emotional stress through non-judgmental tones, their verbose outputs increase cognitive load \cite{2025-Alami}. \yzz{Crucially, LLM-assisted reviews may fail to foster collective accountability, as the lack of social validation hinders the shared ownership emergent in human interactions \cite{2025-Alami-TOSEM}.} This underscores the need for hybrid workflows that retain human contextual oversight against bias and vulnerabilities \cite{2025-Salem}. Developers often view LLMs as useful for reference but react negatively when tasks are fully outsourced due to bugs and unhelpful suggestions \cite{2024-Watanabe}.

\begin{center}
    \begin{tcolorbox}[colback=gray!10!white,colframe=black!75!black,title=\textbf{Summary for Understanding Studies}]
            \begin{enumerate}
                \item \textbf{Quality assurance} requires project-specific customization to address current gaps in detecting security vulnerabilities and reviewing specialized artifacts.
                
                \item \textbf{Process efficiency} depends on optimizing patch size and turnaround time, though challenges like cognitive bottlenecks, information overload, and superficial ``LGTM'' approvals persist.
                
                \item \textbf{Human factors} and demographic biases significantly shape review outcomes, emphasizing that knowledge transfer and psychological safety are as critical as defect detection.
                
                \item \textbf{Human-AI collaboration} driven by LLMs offers powerful automation but necessitates alignment to mitigate over-reliance, trust issues, and the erosion of collective ownership.
            \end{enumerate}
    \end{tcolorbox}
\end{center}

\section{A Roadmap for the Future of MCR}
\label{sec:roadmap}

\begin{table*}[htbp]
    \centering
    \caption{SWOT Analysis Matrix of the Current MCR Research Landscape.}
    \label{tab:swot_matrix}
    \renewcommand{\arraystretch}{1.5}
    \small
    
    \begin{tabularx}{\textwidth}{X!{\vrule width 2pt}X}
        
        \textbf{\large Strengths (Internal)} & \textbf{\large Weaknesses (Internal)} \\
        \textit{Technological Capabilities} & \textit{Technological Limitations} \\
        \vspace{-0.5em}
        \begin{itemize}[leftmargin=*, nosep, topsep=4pt, itemsep=4pt]
            \item \textbf{Unified Frameworks:} Emergence of multi-task architectures that simultaneously handle quality estimation, comment generation, and code refinement \cite{2022-Li, 2023-Lu, 2024-Vijayvergiya}.
            \item \textbf{Generative Proficiency:} LLMs demonstrate superior capability in synthesizing code patches and review comments compared to traditional deep learning models, enabling zero-shot automation \cite{2024-Guo, 2024-Lu}.
            \item \textbf{Emerging Agency:} Autonomous multi-agent systems are beginning to simulate role-based collaboration, utilizing debate and reflection to reduce single-model errors \cite{ren2025hydra, 2024-Tang}.
        \end{itemize}
        & 
        \vspace{-0.5em}
        \begin{itemize}[leftmargin=*, nosep, topsep=4pt, itemsep=4pt]
            \item \textbf{Context Gap:} Models struggle to access and reason about project-specific history, issue trackers, and design documentation, often resulting in superficial suggestions \cite{2023-Lu, 2025-Case}.
            \item \textbf{Hallucination \& Trust:} High false-positive rates and superficial approvals increase verification overhead and erode developer trust \cite{2025-Cihan, 2025-Peng}.
            \item \textbf{Misaligned Metrics:} Reliance on similarity-based metrics fails to capture the constructiveness, logic, and educational value of review feedback \cite{2023-Zhou, 2024-Tufano}.

        \end{itemize}
        \\ 
        \noalign{\hrule height 2pt}
        
        \textbf{\large Opportunities (External)} & \textbf{\large Threats (External)} \\
        \textit{Socio-Technical Environment Potential} & \textit{Socio-Technical Environment Risks} \\
        \vspace{-0.5em}
        \begin{itemize}[leftmargin=*, nosep, topsep=4pt, itemsep=4pt]
            \item \textbf{Data Flywheels:} Leveraging industrial feedback loops and cleaning mechanisms (e.g., CuRev) to continuously refine model precision via high-fidelity data \cite{2025-Sun, 2025-Sghaier}.
            \item \textbf{Specialized Domains:} High potential for tailored applications in vertical domains such as security auditing, education mentorship, and green software engineering \cite{takieldeen2025unlocking, 2025-Ardic, sharanarthi2025adaptive}.
            \item \textbf{Process Integration:} Embedding AI into CI/CD pipelines and IDEs to provide real-time, low-friction assistance rather than post-hoc critiques \cite{2023-Palvannan}.
        \end{itemize}
        & 
        \vspace{-0.5em}
        \begin{itemize}[leftmargin=*, nosep, topsep=4pt, itemsep=4pt]
            \item \textbf{Erosion of Ownership:} Over-reliance on automation may diminish collective responsibility, reciprocity, and social validation within development teams \cite{2025-Alami-TOSEM}.
            \item \textbf{Deskilling:} Junior developers risk losing critical learning opportunities derived from manual review and debugging processes \cite{2025-Tseng}.
            \item \textbf{Bias Amplification:} LLMs may propagate or exaggerate historical biases (e.g., gender bias, toxic tone) present in training corpora, harming community health \cite{2024-Rahman, 2023-Murphy-Hill}.
        \end{itemize}
        \\
    \end{tabularx}
\end{table*}

Based on the comprehensive review of improvement techniques and understanding studies, this section aims to systematically delineate the future research trajectory of MCR. While recent advancements, particularly in LLMs, have revolutionized the technical landscape of code review, a significant misalignment persists between these computational capabilities and the complex socio-technical realities of industrial workflows. To rigorously diagnose this disparity and identify high-value research directions, we employ a SWOT (Strengths, Weaknesses, Opportunities, and Threats) analysis framework. This strategic diagnosis serves as the foundation for proposing three pivotal pillars for future research: Context-Aware Proactivity, Value-Driven Evaluation, and Human-Centric Symbiosis.

\subsection{SWOT Analysis of the MCR Landscape}
SWOT analysis is a strategic planning technique used to identify internal capabilities (Strengths and Weaknesses) and external environmental factors (Opportunities and Threats). In the context of MCR research, we map ``Internal Factors'' to the technological capabilities of current automation models, and ``External Factors'' to the broader socio-technical environment involving developers, organizations, and open-source communities. The detailed matrix is presented in Table \ref{tab:swot_matrix}, while the in-depth analysis of these dimensions follows below.

\subsubsection{Internal Factors: Technological Capabilities vs. Limitations}
The internal dimension evaluates the inherent properties of current state-of-the-art approaches.

\textbf{Strengths.} Unlike previous rule-based methods, \textit{unified frameworks} enable deep learning models that can simultaneously perform defect detection, comment generation, and code refinement without task-specific architecture modifications \cite{2022-Li, 2023-Lu}. Complementing the advancements of training approaches, the primary strength of the current generation of MCR tools lies in LLMs' \textit{generative proficiency}, which demonstrates an unprecedented ability to understand and even synthesize code changes across multiple languages \cite{2024-Guo, 2025-Ramesh}. Most recently, the emergence of \textit{autonomous agency} marks a significant leap, where multi-agent systems can simulate collaboration to self-correct and refine outputs \cite{2024-Tang, ren2025hydra}.

\textbf{Weaknesses.} Despite these strengths, technological limitations remain severe. The most critical bottleneck is the \textit{Context Gap}. Current models typically operate on local code snippets or file-level diffs, lacking the capacity to reason about the broader project history, issue tracking systems, or architectural design documents \cite{2023-Lu, 2025-Case}. Furthermore, the issue of \textit{Hallucination and Trust} persists. Models frequently exhibit high false-positive rates or generate superficial approvals, such as ``LGTM smells'', that appear plausible but mask deeper issues—thereby increasing verification overhead and eroding developer trust \cite{2025-Cihan, 2025-Peng}. Additionally, the field suffers from \textit{Misaligned Metrics}. The reliance on n-gram-based evaluation metrics (e.g., BLEU \cite{BLEU}) fails to capture the logic, constructiveness, and educational value of review feedback \cite{2023-Zhou}, resulting in a ``blind spot'' where improvements in metrics do not translate to practical utility.

\subsubsection{External Factors: Socio-Technical Potential vs. Risks}
The external dimension examines how these technologies interact with the human and organizational environment.

\textbf{Opportunities.} The evolving software ecosystem offers significant opportunities for integration. The concept of \textit{Data Flywheels} in industrial settings allows for the continuous refinement of models using high-fidelity feedback loops from professional developers \cite{2025-Sun, 2025-Sghaier}. Moreover, there is immense potential in \textit{Specialized Domains}. By tailoring general-purpose models to vertical applications—such as security auditing, education mentorship, and green software engineering—researchers can address high-stakes problems that generic tools overlook \cite{takieldeen2025unlocking, 2025-Ardic}. Finally, mirroring the paradigm shift in code generation driven by tools like Cursor\footnote{\url{https://cursor.com/}} and Claude-Code\footnote{\url{https://claude.com/product/claude-code}}, a significant opportunity lies in \textit{Process Integration}. The field currently lacks a ``killer app'' that seamlessly embeds intelligence into the review workflow, moving beyond passive bots to provide real-time, context-aware assistance directly within the developer's environment \cite{2023-Palvannan}.

\textbf{Threats.} Conversely, the deployment of these technologies introduces profound socio-technical risks. The most pressing threat is the \textit{Erosion of Ownership}. Over-reliance on automation may diminish the sense of collective responsibility and social reciprocity that binds development teams together \cite{2025-Alami-TOSEM}. Furthermore, the phenomenon of \textit{Deskilling} poses a long-term risk to workforce sustainability, as junior developers may lose critical learning opportunities derived from the manual review process \cite{2025-Tseng}. Finally, without careful governance, AI models risk \textit{Bias Amplification}, potentially propagating historical gender biases or toxic communication patterns present in training corpora \cite{2024-Rahman}.

\subsection{Future Direction}
As illustrated in Table \ref{tab:swot_matrix}, the SWOT analysis reveals a clear imperative: technological prowess (Strengths) alone cannot overcome the contextual limitations (Weaknesses) or mitigate socio-technical risks (Threats). 
the current state of MCR research is characterized by a tension between burgeoning generative capabilities and persistent contextual limitations.
Internally, while LLMs demonstrate \textit{Generative Proficiency} and the ability to operate within \textit{Unified Frameworks} \cite{2022-Li, 2024-Guo}, they suffer from a \textit{Context Gap}, often failing to access project-specific history or design constraints \cite{2023-Lu, 2025-Case}.
Externally, the field faces the ``Opportunity'' to leverage industrial data flywheels \cite{2025-Sun} and penetrate specialized domains like security and education \cite{takieldeen2025unlocking, 2025-Ardic}. However, these advances bring critical ``Threats,'' most notably the potential erosion of collective ownership \cite{2025-Alami-TOSEM} and the risk of deskilling junior developers \cite{2025-Tseng}.
Future research must therefore pivot from purely pursuing model accuracy to addressing these systemic challenges. Consequently, we propose three paradigm shifts to guide the next phase of MCR research.

\subsubsection{Paradigm Shifts I: From Passive Assistants to Proactive Collaborators}
To overcome the \textit{Context Gap} and \textit{Hallucination} weaknesses identified in the SWOT analysis, future technical research must move beyond generic code understanding toward systems that possess deep project awareness and autonomous agency.

\begin{itemize}
    \item \textbf{Repository-Specific Contextualization via Long-term Memory.}
    Current models largely treat code changes in isolation or within a limited context window, ignoring the distinct coding styles \cite{2018-Ueda}, conventions \cite{2020-Han}, and review workflows \cite{2019-Fatima2} unique to each repository. To address this, future frameworks must integrate Retrieval-Augmented Generation (RAG) to establish a Long-term Project Memory. This involves not only retrieving similar past commits but also linking code changes to non-code artifacts, such as issue trackers, design documents, and historical review discussions \cite{2025-Jaoua, sharanarthi2025multi}. By grounding generation in the specific semantic context of the repository, models can transition from offering generic syntax corrections to providing architectural-level insights that strictly align with the project's evolution.
    \item \textbf{Multi-Agent Collaboration Ecosystems.}
    Given the complexity of review tasks, ranging from security auditing to style checking, a single monolithic model often struggles to balance competing objectives. The future lies in ``Multi-Agent Systems'' where specialized agents (e.g., a Security Agent, a Refactoring Agent, and a QA Agent) collaborate. These agents should transcend passive responsiveness by proactively offering insights and suggestions across various stages of the review lifecycle. Furthermore, to facilitate repository-specific generation, each agent must be equipped with long-term memory capabilities, enabling them to align their specialized advice with the project's historical evolution. As demonstrated by early works like HYDRA-REVIEWER \cite{ren2025hydra} and role-based simulations \cite{2024-Tang}, such agents can employ debate and reflection mechanisms to cross-verify suggestions, thereby mitigating hallucinations and improving the precision of feedback.
    \item \textbf{Dynamic Knowledge Evolution.}
    While LLMs and pre-trained models (PTMs) possess robust generalized capabilities, they inevitably face a temporal mismatch: their static parametric knowledge rapidly becomes obsolete amidst the high-velocity iteration of modern software projects. To bridge this gap, future systems must implement \textit{Data Flywheels} that continuously capture real-time feedback from human reviewers (e.g., accepted or rejected comments). Adaptive mechanisms, triggered by metrics such as the Outdated Rate \cite{2025-Sun}, should be employed to ensure synchronization with the living codebase. Crucially, this evolving data serves a dual purpose in empowering proactive agents. On the one hand, it acts as a dynamic \textit{knowledge base} for LLMs, supplementing their reasoning with the latest project-specific context to ensure accurate, high-level review assistance. On the other hand, it provides high-quality training corpora for the \textit{rapid iteration} of specialized PTMs, enabling them to adapt quickly to latency-sensitive, interactive scenarios—such as real-time code change improvement—where immediate responsiveness to developer intent is paramount.

\end{itemize}

Building upon the proactive collaborators paradigm, the ultimate evolution of MCR tooling lies in \textbf{Contextual Ubiquity}. We envision a transition from traditional asynchronous, web-centric pull request interfaces toward \textbf{real-time, IDE-native agents} that permeate every stage of the code review lifecycle. By continuously evolving through the ingestion of historical project data, these agents facilitate a \textbf{``Pre-emptive Review''} model. In this paradigm, AI provides immediate, context-aware guidance during the initial authoring phase, significantly alleviating the downstream burden on human reviewers. This shifts the review process from a static, post-hoc gatekeeping task into a dynamic, interactive dialogue. Crucially, such pervasive intelligence serves as a \textbf{catalyst for knowledge dissemination}, fostering deeper mutual understanding and streamlined expertise transfer between authors and reviewers.

\subsubsection{Paradigm Shifts II: From Accuracy Metrics to Value-Driven Evaluation}
The SWOT analysis highlights \textit{Misaligned Metrics} as a critical weakness. To ensure that AI tools genuinely enhance MCR rather than merely increasing noise, the evaluation paradigm must shift from measuring textual similarity to assessing practical utility and cognitive impact.

\begin{itemize}
    \item \textbf{Beyond Similarity: Utility, Acceptability, and Impact.}
    Traditional metrics, including n-gram overlaps (e.g., BLEU \cite{BLEU}, ROUGE \cite{chin2004rouge}) and edit similarity, fundamentally rely on measuring the textual proximity between generated outputs and historical ground truth. However, as automated MCR models evolve, their generated review suggestions and code refinements may substantively surpass the quality of the original human references. Consequently, the prevailing similarity-based paradigm is becoming obsolete, as ``deviation'' from the ground truth no longer implies ``error'' but potentially signifies ``improvement.'' Future benchmarks must therefore prioritize Utility \cite{2023-Ahmed}, Acceptability \cite{2023-Yang}, and the holistic Impact on repository quality (e.g., future bug reduction and maintainability improvements). This shift necessitates the development of ``LLM-as-a-Judge'' frameworks and high-dimensional semantic evaluation protocols capable of assessing whether a comment provides actionable advice, correctly identifies logic errors, or offers valuable rationale, rather than simply matching a reference string.

    \item \textbf{Measuring Cognitive Load and Verification Cost.}
    Models that appear high-performing on static benchmarks should be deemed \textit{failures} if they generate ``LGTM smells'' or subtle bugs that require extensive human verification. Critically, such subtle inaccuracies often entail more severe consequences than overtly erroneous predictions, as they induce a false sense of security and demand disproportionate cognitive effort to identify \cite{2024-Gon, 2025-Cihan}. Therefore, evaluation must incorporate ``Cognitive Impact Metrics,'' such as Verification Latency (the time taken for a human to validate an AI suggestion) and the downstream consequences of adopting such suggestions. A successful tool must demonstrate a net reduction in total review time while maintaining robust defect detection rates—treating correctness as a necessary baseline rather than the sole optimization target—explicitly accounting for the cost of reviewing the AI's output itself \cite{2025-Tufano}.

    \item \textbf{Benchmarking for Real-world Complexity.}
    High-quality benchmarks serve as a compass for a research field's advancement \cite{swebench}. However, existing datasets predominantly feature isolated, atomic changes that diverge significantly from industrial reality. To mirror real-world challenges, future benchmarks must simulate complex, multi-file pull requests and iterative, conversational feedback loops. While recent initiatives like SWR-Bench \cite{2025-Zeng} have begun to address these gaps, they remain limited in scope and extensibility. Crucially, evaluations must assess an AI's ability to maintain context across multi-turn interactions and manage ``Merge Request deviations'' \cite{2025-Kansab}. Furthermore, significant version updates are rarely encapsulated in a single PR but rather involve extensive divergence across repository branches. Consequently, benchmarks must expand to include cross-branch comparisons and the review of long-lived feature branches, moving beyond the traditional single-PR paradigm.

\end{itemize}

\subsubsection{Paradigm Shifts III: From Automation to Human-Centric Symbiosis}
Finally, addressing the socio-technical threats of \textit{Erosion of Ownership}, \textit{Deskilling}, and \textit{Bias Amplification} requires re-imagining the role of AI not as a replacement for human reviewers, but as a collaborative partner that actively supports social, educational, and ethical goals.

\begin{itemize}
    \item \textbf{Governance for Collective Ownership.}
    To mitigate the risk of reduced social responsibility \cite{2025-Alami-TOSEM}, robust governance mechanisms must be integrated into AI workflows. This includes ``Human-in-the-Loop'' protocols that mandate explicit human sign-off for critical architectural changes and safety gates, thereby preventing the uncritical acceptance or automation bias of AI-generated code. Moreover, AI assistance should transcend the generation of isolated artifacts for single tasks. By intelligently recommending optimal reviewers \cite{2025-Rigby} and actively surfacing necessary contextual references, AI acts as a facilitator that cultivates a constructive review atmosphere, reinforcing the social fabric of the team rather than isolating developers behind automated interfaces.
    
    \item \textbf{AI as a Mentor.}
    To prevent the loss of critical learning opportunities for junior developers \cite{2025-Tseng}, AI tools must transcend simple error correction and embrace pedagogical goals. Therefore, ``Mentorship-oriented AI'' should provide detailed explanations (Rationale) for its suggestions \cite{2025-Widyasari} and guide novices through the reasoning process via Socratic dialogue. This approach transforms code review from a mere gatekeeping activity into a continuous learning environment, ensuring that automation augments rather than atrophies human expertise. In parallel, the proliferation of repository-level code understanding systems, such as DeepWiki\footnote{\url{https://deepwiki.com/}}, is poised to shoulder the broader responsibility of knowledge sharing. By offloading general context dissemination to these specialized tools, the review process itself is liberated to focus on high-value cognitive skill transfer.

    \item \textbf{Ethical AI for Inclusive Communities.}
    Recognizing that MCR is a deeply social process, future research must proactively address the threat of \textit{Bias Amplification}. The automation models trained on historical data risk propagating or even exaggerating systemic biases, such as gender-based exclusion \cite{2023-Murphy-Hill} or toxic communication patterns \cite{2024-Rahman}. Inspired by recent advancements in aligned AI (e.g., Anthropic's Constitutional AI which emphasizes helpfulness and harmlessness \cite{2022-HarmlessnessAI}), AI assistants must incorporate active ``Detoxification'' and ``Bias Detection'' layers. Beyond simply filtering toxic language, these systems should be designed to foster psychological safety \cite{2024-Murphy-Hill} by nudging reviewers toward more inclusive and constructive phrasing. This ensures that the efficiency gains of automation do not come at the cost of equity and community health.
\end{itemize}

In summary, the future success of MCR depends on a reciprocal relationship between technical innovation and empirical inquiry. \textbf{Understanding studies} serve as the ``compass,'' identifying human-centric requirements—such as the specific information needs of reviewers \cite{2018-Ebert} and the socio-technical impacts of bias \cite{2022-Gunawardena}—which define the design space for new tools. Conversely, \textbf{improvement techniques} act as the ``engine,'' providing the capabilities to address these bottlenecks. As new AI-collaborative tools are deployed, they inevitably create new human behaviors, such as over-reliance or reduced visual attention \cite{2022-Madi}. This creates a continuous feedback loop: technical improvements necessitate new understanding studies, which in turn refine the requirements for the next generation of automation. This co-evolution ensures that MCR intelligence remains not only powerful but also practically useful and socially responsible.

\section{Conclusion}
\label{sec-conclusion}

This paper presents a comprehensive roadmap for Modern Code Review (MCR), establishing a unified framework that bridges over a decade of improvement techniques and understanding studies (2013--2025). By synthesizing 327 studies and performing a strategic SWOT analysis, we provide a rigorous diagnosis of the current MCR landscape, uncovering a profound misalignment between burgeoning AI capabilities and industrial realities. We identify that while Generative AI offers unprecedented proficiency, the field is hindered by critical technical barriers---namely the repository context gap, stochastic hallucinations, and misaligned evaluation metrics---alongside systemic socio-technical threats including the erosion of collective ownership, junior developer deskilling, and algorithmic bias amplification. To navigate this transformative junction, we propose three foundational paradigm shifts: transitioning toward Context-Aware Proactivity through project-specific long-term memory; adopting Value-Driven Evaluation centered on actionable utility and cognitive impact; and cultivating Human-Centric Symbiosis where AI serves as a mentor and architectural guardian. Ultimately, this roadmap provides a cohesive vision for the research community to transform MCR from a resource-intensive bottleneck into an intelligent, inclusive, and strategic asset that harmonizes computational power with the social essence of human-driven code inspection.

\begin{acks}
This research is supported by National Natural Science Foundation of China under project (No. 62472126) and CCF-Huawei Populus Grove Fund.
\end{acks}

\bibliographystyle{ACM-Reference-Format}
\bibliography{ref}

\end{document}